\documentclass{article}

\usepackage{microtype}
\usepackage{graphicx}
\usepackage{subcaption}
\usepackage{enumitem}

\usepackage{booktabs} 
\newcommand{\ba}{\textsc{BibAgent}}
\newcommand{\mb}{\textsc{MisciteBench}}

\usepackage{booktabs}
\usepackage{multirow}
\usepackage{siunitx}

\usepackage{xcolor}
\usepackage{listings}

\usepackage[most]{tcolorbox}
\tcbuselibrary{listings} 

\lstdefinestyle{paperCode}{
  basicstyle=\ttfamily\small,
  breaklines=true,
  showstringspaces=false,
  tabsize=2
}

\newtcblisting{codebox}[2][]{%
  listing engine=listings,
  listing options={style=paperCode, language=#2},
  listing only,
  colback=black!2,
  colframe=black!20,
  arc=2mm,
  boxrule=0.5pt,
  left=6pt,right=6pt,top=6pt,bottom=6pt,
  enhanced jigsaw,
  breakable,
  title={#1},
  fonttitle=\bfseries,
}

\newlength{\OrgIconH}
\newcommand{\orgicon}[2][1.0]{%
  \raisebox{-0.15\height}{\scalebox{#1}{\includegraphics[height=\OrgIconH]{figs/#2}}}%
}

\newcommand{\OpenAIIcon}{\orgicon[1.50]{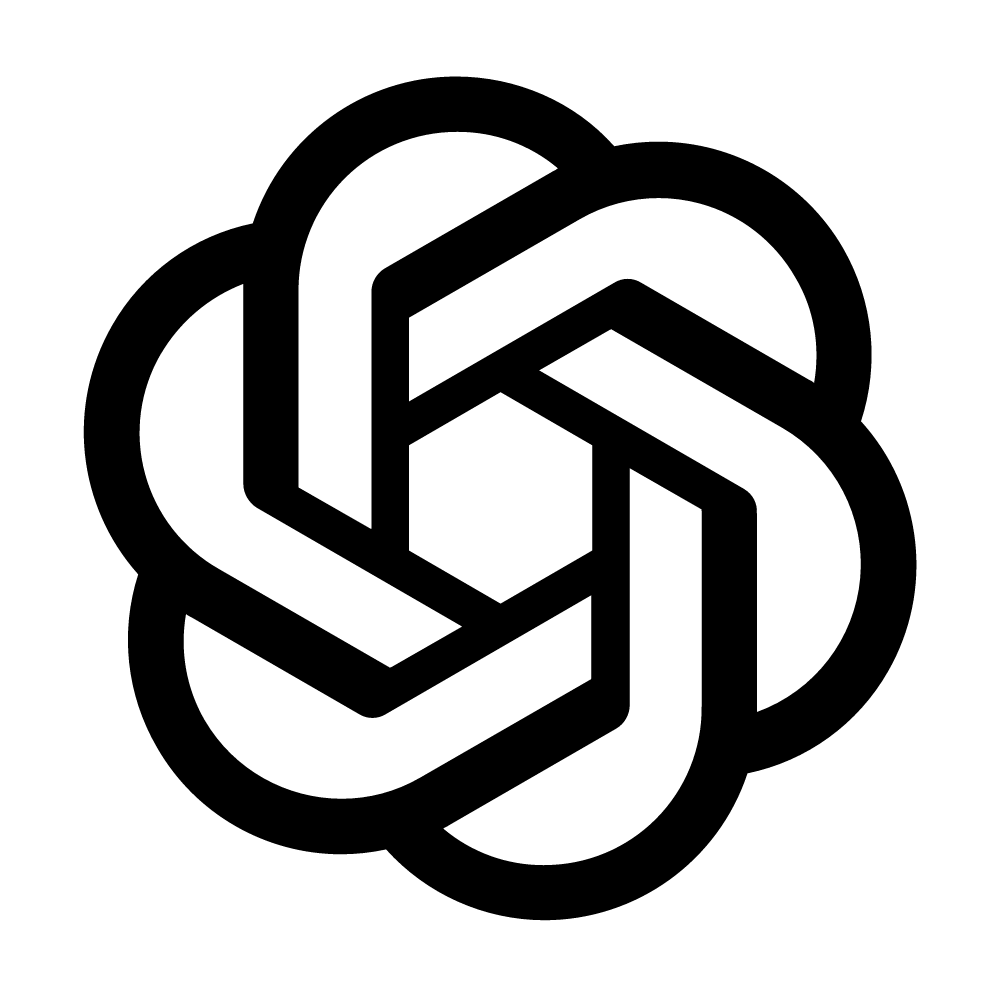}}
\newcommand{\AnthropicIcon}{\orgicon[1.50]{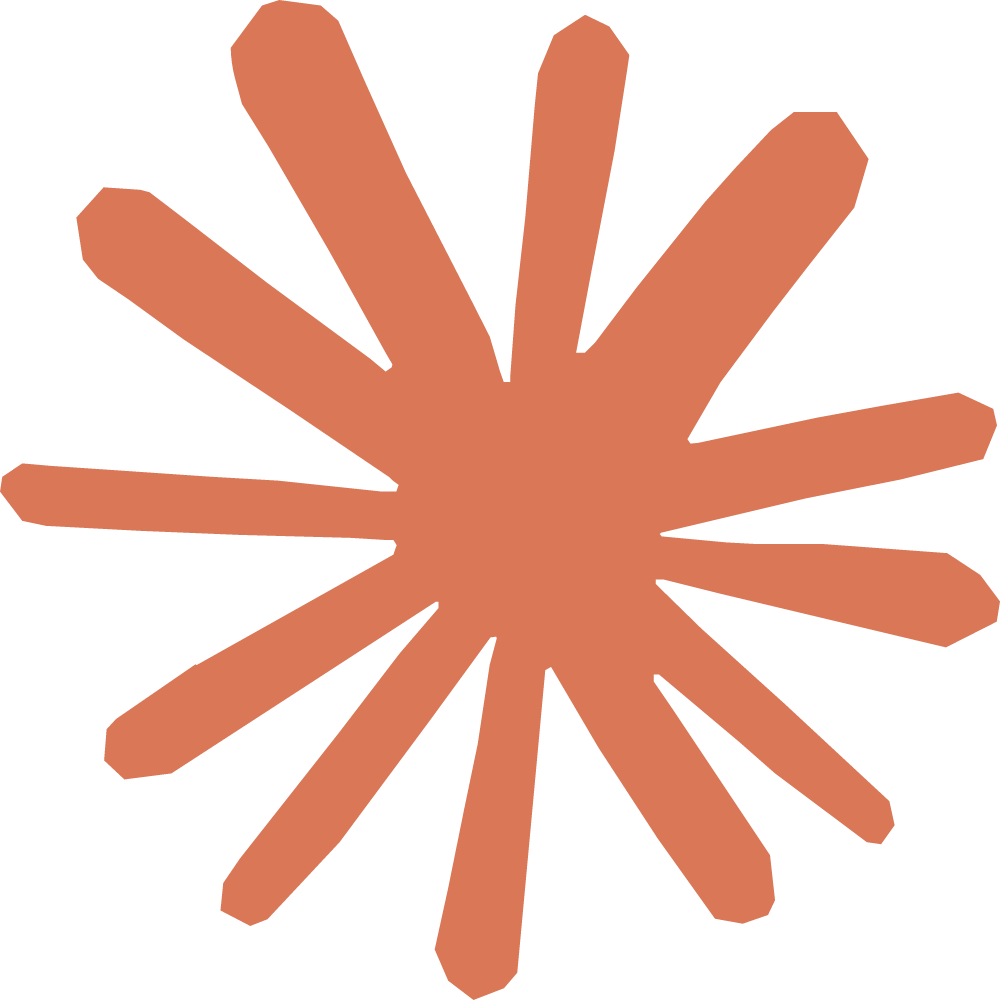}}
\newcommand{\GeminiIcon}{\orgicon[1.50]{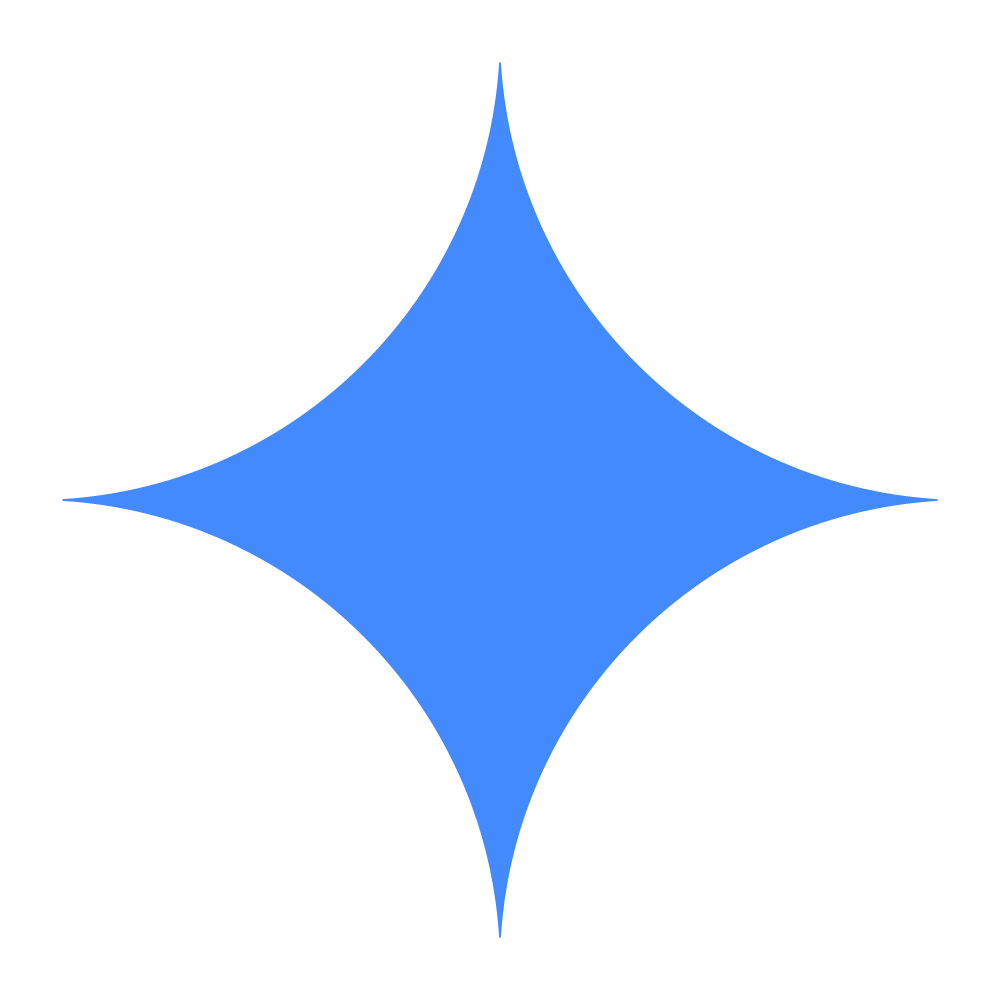}}
\newcommand{\NVIDIAIcon}{\orgicon[1.5]{}}
\newcommand{\MetaIcon}{\orgicon[1.2]{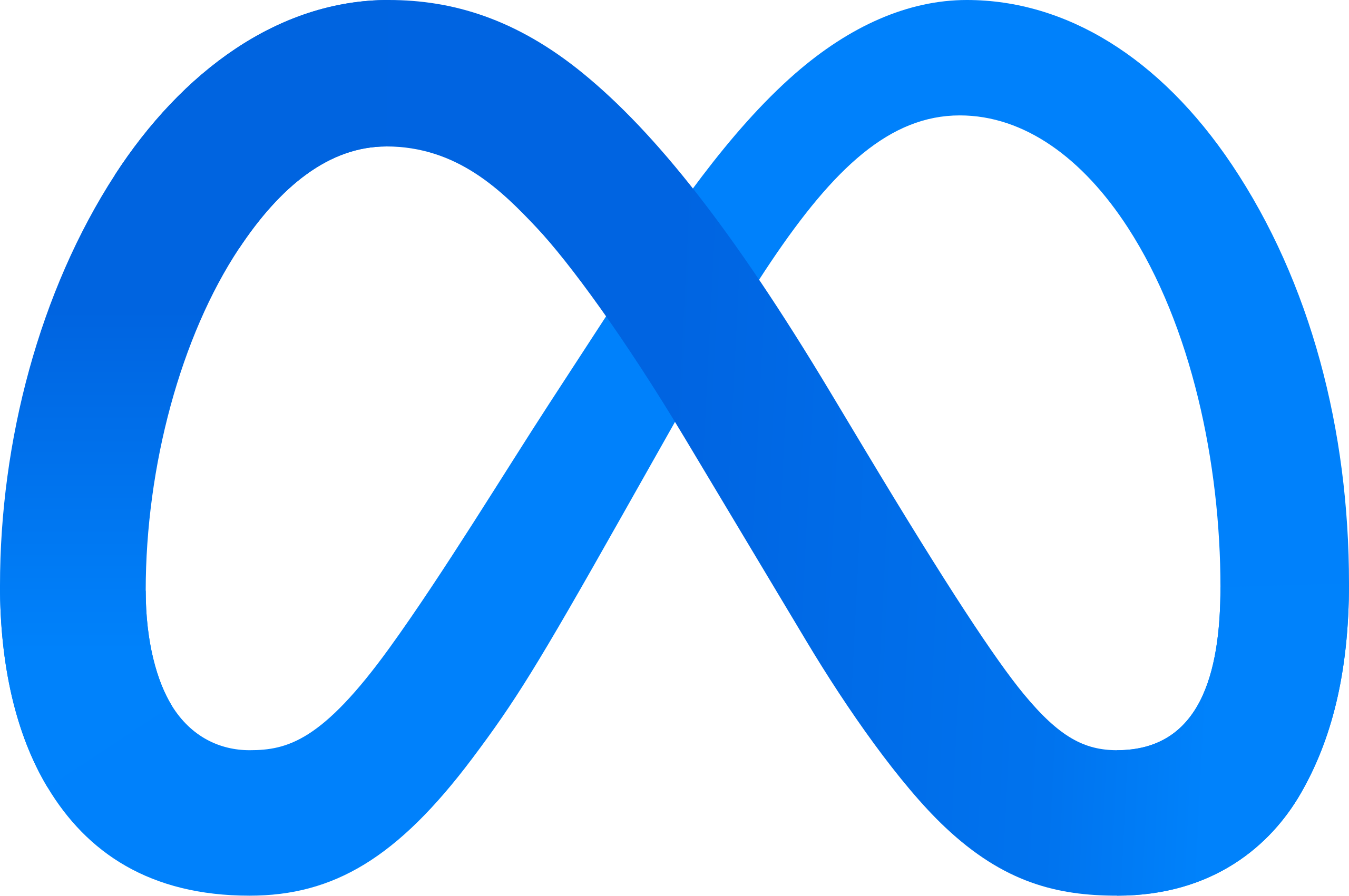}}
\newcommand{\QwenIcon}{\orgicon[1.50]{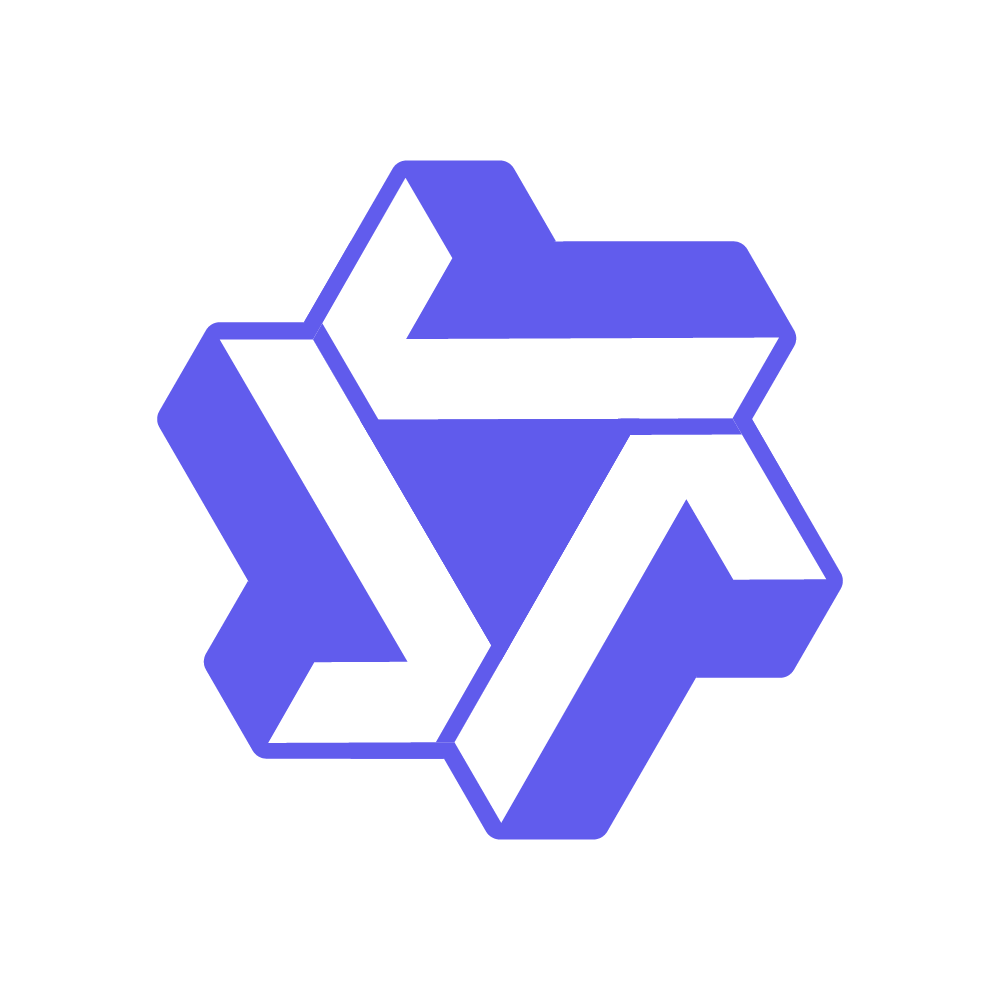}}
\newcommand{\MistralIcon}{\orgicon[1.50]{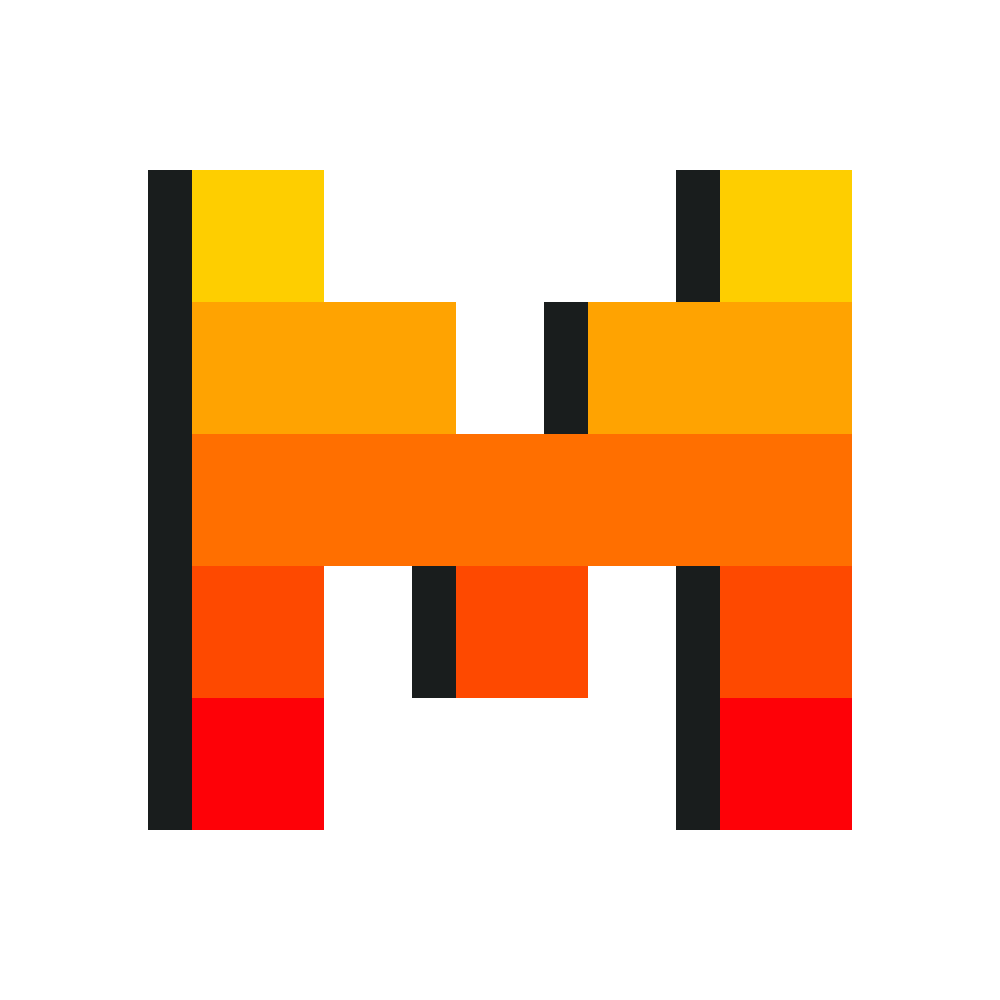}}
\newcommand{\DeepSeekIcon}{\orgicon[1.50]{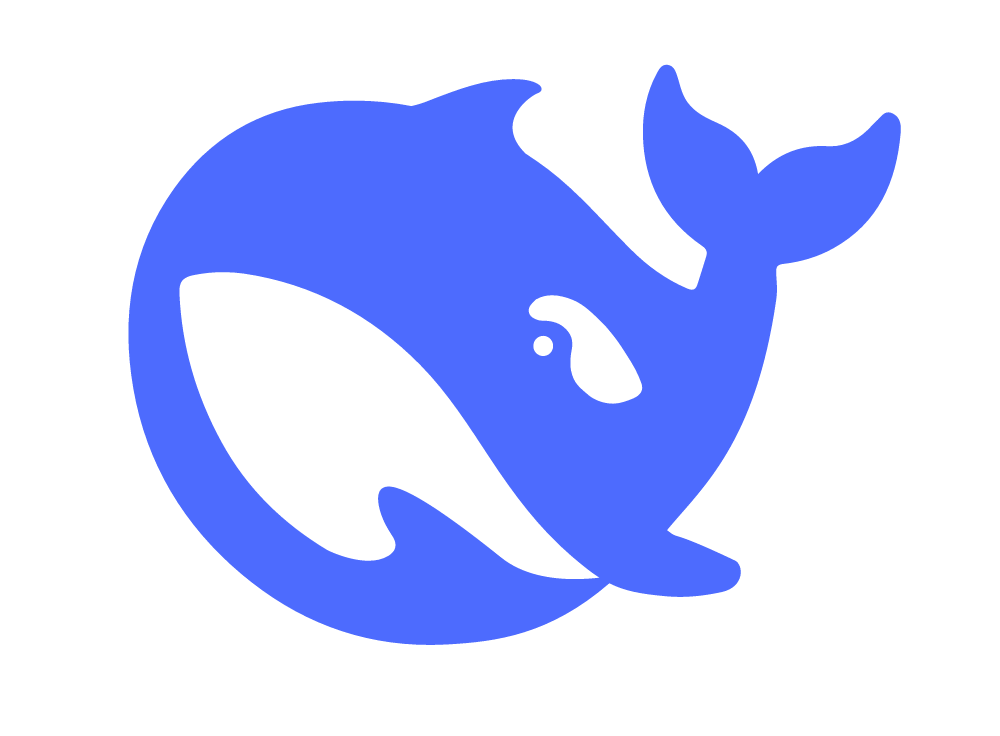}}

\usepackage{hyperref}

\usepackage[preprint]{icml2026}

\usepackage{amsmath}
\usepackage{amssymb}
\usepackage{mathtools}
\usepackage{amsthm}
\usepackage{float}
\usepackage{tabularx}
\usepackage{multirow}
\usepackage{graphicx}

\usepackage[capitalize,noabbrev]{cleveref}

\theoremstyle{plain}

\theoremstyle{definition}

\theoremstyle{remark}

\usepackage[textsize=tiny]{todonotes}

\icmltitlerunning{BibAgent: An Agentic Framework for Traceable Miscitation Detection in Scientific Literature}

\begin{document}


\twocolumn[
  \icmltitle{\textsc{\ba:} An Agentic Framework for Traceable Miscitation \\Detection in Scientific Literature}



  \icmlsetsymbol{equal}{*}

  \begin{icmlauthorlist}
    \icmlauthor{Peiran Li}{a,b}
    \icmlauthor{Fangzhou Lin}{a,c}
    \icmlauthor{Shuo Xing}{a}
    \icmlauthor{Xiang Zheng}{b}
    \icmlauthor{Xi Hong}{b} \\ 
    \icmlauthor{Siyuan Yang}{a}
    \icmlauthor{Jiashuo Sun}{d}
    \icmlauthor{Zhengzhong Tu}{a}
    \icmlauthor{Chaoqun Ni}{b}
  \end{icmlauthorlist}

  \icmlaffiliation{a}{Texas A\&M University}  
  \icmlaffiliation{b}{University of Wisconsin-Madison}
  \icmlaffiliation{c}{Worcester Polytechnic Institute}  
  \icmlaffiliation{d}{University of Illinois Urbana-Champaign}    
 

  \icmlcorrespondingauthor{Peiran Li}{lipeiran@tamu.edu}
  \icmlcorrespondingauthor{Chaoqun Ni}{chaoqun.ni@wisc.edu}

  \icmlkeywords{Machine Learning, ICML}

  \vskip 0.3in

{

\begin{center}
\vspace{-16px}
    \centering
    \captionsetup{type=figure}
    \includegraphics[width=\textwidth]
    {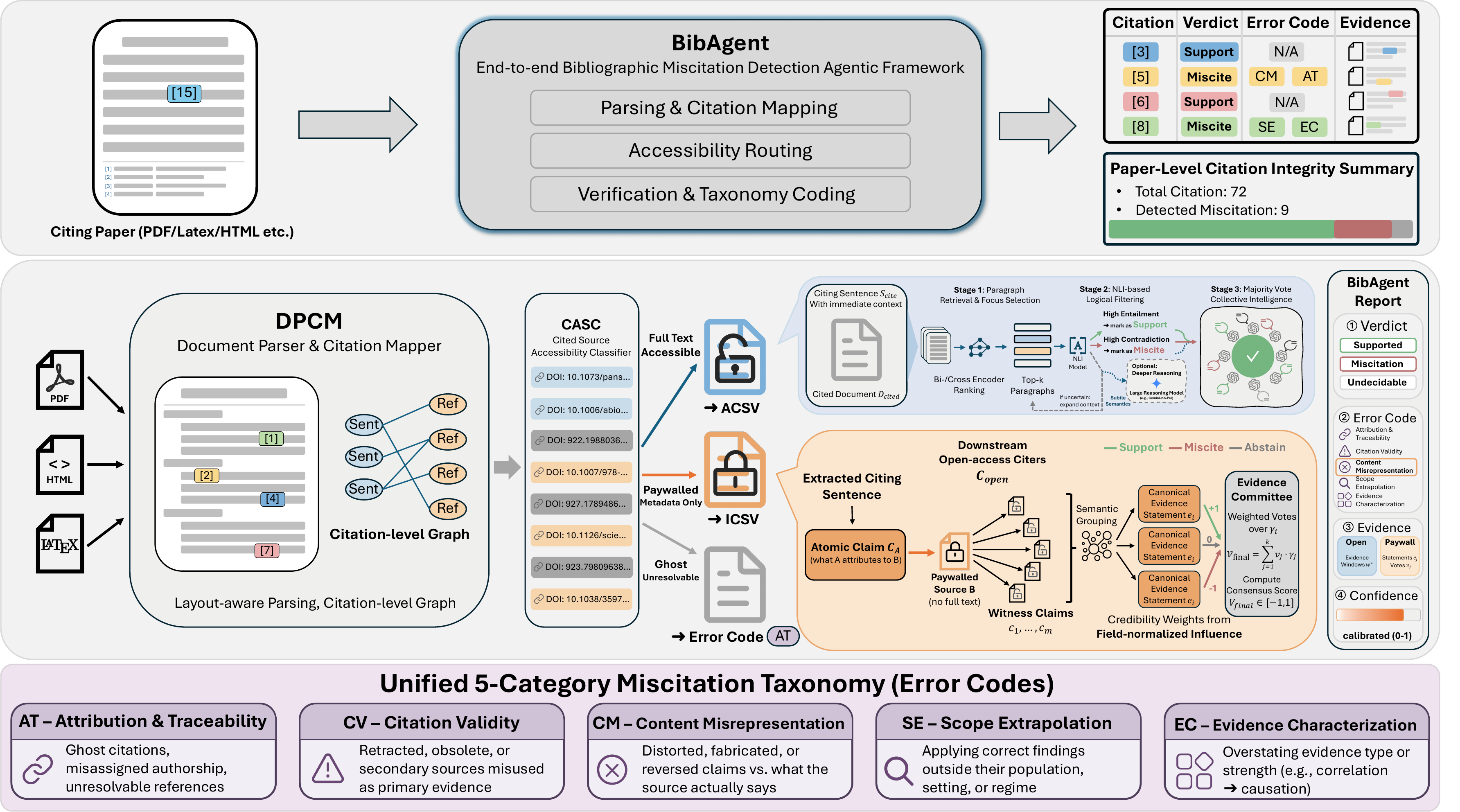}
    \captionof{figure}{\protect{\textbf{\ba~: Traceable Miscitation Detection for Accessible and Paywalled Scientific Literature.}} End-to-end pipeline: DPCM parses papers and maps in-text citations; CSAC routes each reference to ACSV (full-text evidence localization) or ICSV (Evidence Committee over downstream citers) under paywalls; the system emits per-citation verdicts, calibrated confidence, and taxonomy-aligned error codes (AT/CV/CM/SE/EC) with traceable evidence.}
    \label{fig:teaser}
\end{center}%
\vskip 0.1in

}

]





\printAffiliationsAndNotice{}  

\begin{abstract}
  Citations are the bedrock of scientific authority, yet their integrity is often compromised by widespread miscitations---ranging from nuanced distortions to fabricated references. 
  Systematic citation verification is currently unfeasible; manual review cannot scale to modern publishing volumes, while existing automated tools are restricted by abstract-only analysis or small-scale, domain-specific datasets in part due to the ``paywall barrier" of full-text access. We introduce \ba, a scalable, end-to-end agentic framework for traceable citation verification. 
  \ba~integrates retrieval, reasoning, and adaptive evidence aggregation, applying distinct strategies for accessible and paywalled sources via leveraging a novel \emph{Evidence Committee} mechanism that infers citation validity via downstream citation consensus. 
  To support systematic evaluation, we introduce \mb, a massive cross-disciplinary benchmark comprising 6,350 miscitation samples spanning 254 fields. 
  Our experimental results demonstrate that \ba~ outperforms state-of-the-art Large Language Model (LLM) baselines in citation verification accuracy and interpretability, providing scalable, transparent detection of citation misalignments across the scientific literature.  
  
\end{abstract}

\section{Introduction}

Citations are the infrastructure of scientific communication: they determine how claims are justified, how credit is allocated, and how careers and funding decisions are made~\cite{waltman2016review}. In principle, a citation is a verifiable contract between a statement and its evidence. In practice, that contract is frequently broken. Large-scale audits report that \textbf{39\%} of sampled biomedical citations are inaccurate~\cite{sarol2024assessing}, and misquotation rates of \textbf{10\%--25\%} have been documented across disciplines~\cite{de1985accurate,jergas2015quotation,rekdal2014academic}. For instance, clinical guidelines sometimes still cite retracted trials as primary evidence, or LLM-written drafts attribute causal claims to purely correlational studies. We use \textbf{miscitation} to denote failures where a cited source is distorted, fabricated, or invoked outside its valid evidential role. This problem is newly urgent in the era of generative AI: Large Language Models (LLMs) can produce fluent prose accompanied by plausible but hallucinated or misused references at scale~\cite{walters2023fabrication,liang2024mapping,rao2025detecting}. As a result, scientific claims increasingly risk being backed by references that are unread, misrepresented, or even non-existent, undermining both the integrity of the scholarly record and the fairness of evaluation.

Manual citation checking does not scale. Even expert reviewers must cross-read long papers, resolve context and intent, and judge subtle distinctions (e.g., association vs.\ causation, narrow vs.\ broad scope). With modern papers citing more than 45 references on average~\cite{dai2021literary}, exhaustive verification at review or publication time is infeasible. This has motivated automated approaches, including citation-function and citation-sentiment classifiers~\cite{teufel2006automatic,athar-teufel-2012-detection,kilicoglu2019confirm} and claim-verification systems that judge whether short natural-language claims are supported by abstracts or retrieved snippets~\cite{SciFact,li2021paragraphlevelmultitasklearningmodel,wadden-etal-2022-multivers,liu2024retrieval}. More recent work targets citation errors directly~\cite{sarol2024assessing,zhang2024detectingreferenceerrorsscientific}. However, despite rapid progress in LLM-based reasoning, the field still lacks a solution that is simultaneously \emph{taxonomy-aware} (distinguishing \emph{how} a citation fails), \emph{paywall-robust} (operating when full text is inaccessible), and \emph{auditable} (providing traceable evidence for each decision) across disciplines and deployment settings.

A key reason is that citation verification operates in \emph{two regimes} that most prior systems conflate. In the \textbf{accessible} regime, the bottleneck is not just deciding support vs.\ refutation, but \emph{localizing decisive evidence} in long, structured documents and diagnosing fine-grained miscitation types (e.g., scope extrapolation or overstated evidence strength) that cannot be reduced to a binary label. In the \textbf{inaccessible} regime, full text may be behind paywalls or otherwise unavailable; treating these cases as uniform abstention creates a systematic blind spot precisely where verification is most needed. Worse, LLM-based verifiers can exhibit sycophancy~\cite{wei2024simplesyntheticdatareduces}: instead of challenging a citing statement against evidence, they may echo its framing, producing confident but ungrounded justifications. Combined with scarce cross-field labeled data and strong disciplinary heterogeneity, these issues have confined existing methods to narrow domains or unrealistic access assumptions, leaving the community without a general, traceable mechanism for maintaining citation integrity in real-world settings.

We start from the view that miscitation detection is fundamentally an \emph{evidence retrieval and reasoning} problem under these two regimes, and that a practical verifier must remain informative even when the cited source is not directly readable. Our central idea is therefore to treat paywalls not as a hard failure mode, but as an alternative evidential setting: when a cited paper is inaccessible, a verifier should reconstruct what the research community collectively attributes to it by aggregating downstream citation contexts, while explicitly modeling noise and abstaining when community evidence is too sparse or inconsistent. In both regimes, decisions should be \emph{traceable}: a reader should be able to see the citing context, the supporting evidence (or reconstructed consensus), and its precise error code.

To operationalize this perspective, we introduce \textbf{\ba}, an agentic framework for bibliographic miscitation detection, together with \textbf{\mb}, a cross-disciplinary benchmark and evaluation suite for miscitation. At the core is a unified \textbf{five-category miscitation taxonomy} that serves as both (i) the label space of \mb\ and (ii) the error-code space predicted by \ba. \ba\ is an end-to-end framework that maps in-text citations to bibliography entries, routes each reference to an \emph{accessible} verification branch (evidence localization and reasoning over full text) or an \emph{inaccessible} verification branch (community-consensus reconstruction from downstream citers), and outputs a single taxonomy-aligned error code for each detected miscitation with traceable evidence. \mb\ instantiates the taxonomy at scale as a contamination-controlled benchmark of \textbf{6{,}350} expert-validated miscitation instances spanning all \textbf{254} Clarivate Journal Citation Reports (JCR) subject categories~\cite{clarivate2025jcr}, designed to stress-test citation reasoning under distribution shift rather than reward memorization.

In summary, our contributions are:
\begin{itemize}[leftmargin=*, topsep=0pt, itemsep=0.5pt, parsep=0pt, partopsep=0pt]
    \item \textbf{A unified miscitation taxonomy and benchmark.} We propose a five-category miscitation taxonomy that provides an operational error-code space for large, heterogeneous corpora, and instantiate it in \mb, a contamination-controlled, cross-disciplinary benchmark of 6{,}350 expert-validated miscitation instances.
    \item \textbf{A paywall-robust, auditable verification agent.} We introduce \ba, an end-to-end bibliographic miscitation detection agent that jointly handles accessible and paywalled sources via specialized verification branches and emits explicit taxonomy-aligned error codes with traceable evidence or reconstructed consensus.
    \item \textbf{Practical miscitation detection at scale.} Across \mb\ and additional corpora, \ba\ improves miscitation detection over strong full-text LLM baselines while substantially reducing computation, and its structured outputs support paper-level citation-integrity summaries with fine-grained, interpretable error codes.
\end{itemize}


\section{Taxonomy \& \mb~Construction}
\label{sec:taxonomy_benchmark}

Automated miscitation detection requires a precise view of \emph{how} citations fail. Prior work largely offers either anecdotal case studies or coarse valid/invalid labels~\cite{pride2017incidental, pride2020authoritative}, and domain-specific schemes in, e.g., clinical medicine or psychology tend to revolve around loosely defined sentiment labels or broad support/contrast tags~\cite{SciFact, wadden-etal-2022-multivers, wadden2021overview, sarol2024assessing}. These typologies are difficult to apply consistently across disciplines and do not provide an operational error-code space for large, heterogeneous corpora. To our knowledge, there is still no unified taxonomy that is both mutually exclusive and collectively exhaustive at the error level and reproducible across fields.

We address this gap by introducing (i) a five-category miscitation taxonomy that underpins our benchmark and agent, and (ii) \mb, a contamination-controlled evaluation suite that instantiates this taxonomy at scale across the scientific literature. The taxonomy gives the error-code space that \ba\ predicts, and \mb\ uses exactly the same code space as its label set. Full category definitions, litmus questions, legacy-label mappings, and the expert annotation protocol are provided in Appendix~\ref{app:taxonomy_annotation} (Table~\ref{tab:miscite_taxonomy_mapping}); here we summarize the components needed to understand the benchmark and subsequent experiments.

\subsection{Five-Category Miscitation Taxonomy}
\label{subsec:taxonomy}

At a high level, our taxonomy partitions miscitation along five orthogonal dimensions that mirror the steps of citation checking. \emph{Citation Validity Errors} capture cases where the source itself is disqualified as scientific evidence (e.g., retracted, obsolete, or a secondary review invoked as if it were primary data). \emph{Content Misrepresentation Errors} arise when the citing text substantively distorts, fabricates, or reverses what the source actually says. \emph{Scope Extrapolation Errors} occur when a correctly understood conclusion is applied outside the population, setting, task, or methodological regime for which it was established. \emph{Evidence Characterization Errors} capture misstatements about the logical type or strength of evidence, such as promoting correlational findings to ``causal'' or describing marginal effects as definitive. Finally, \emph{Attribution \& Traceability Errors} cover broken metadata links---ghost citations, unresolvable references, or misassigned authorship---that prevent a reader from reliably locating and crediting the underlying source.

Each category is defined by a short operational ``litmus question'' and accompanied by positive and negative examples, so that annotators and models can treat miscitation detection as a structured error-coding task rather than an unstructured judgment call. In practice, multiple error types can co-occur in a single citation. To make labels reproducible, we enforce a \emph{Dependency Precedence Rule} that mirrors human verification: one first checks whether a citation can be traced at all, then whether the source is valid, then whether its content is represented faithfully, and only then inspects scope and evidence-strength claims. This rule ensures each miscitation receives exactly one primary category, corresponding to the first point of failure in a realistic citation-checking workflow. Appendix~\ref{app:taxonomy_annotation} gives the complete taxonomy, its mapping to legacy labels, and the expert annotation study validating its usability across disciplines.

\subsection{\mb: A Contamination-Controlled Benchmark}
\label{subsec:miscitebench}

We instantiate this taxonomy in \mb, a large-scale benchmark of 6{,}350 miscitation instances spanning all 254 Clarivate Journal Citation Reports (JCR) subject categories and 21 high-level disciplines (including Agricultural Sciences, Clinical Medicine, Computer Science, Social Sciences, and Visual \& Performing Arts)~\cite{clarivate2025jcr}. Each instance pairs a citing context, a gold evidence span, and a single taxonomy-aligned error code. The design goal is to stress-test citation reasoning under distribution shift, not to reward memorization of a few high-profile works.

Source selection follows a field-normalized, impact-aware protocol (Appendix~\ref{app:source_selection_rationale}). For each JCR subject category, we first identify the 2024 Journal Impact Factor (JIF) leader, then select within that journal the most-cited research article published in 2024--2025. We exclude non–research-article content and low-impact outliers, and manually retrieve full text. To ensure that evaluation cannot be solved from parametric memory, we then apply a \emph{knowledge-blank cleanroom protocol} (Appendix~\ref{app:factcheck}): a panel of strong LLMs receives only metadata (title, authors, venue, year) and is queried with ten forensic questions per candidate paper, each answerable only from the full text. A paper is admitted as a benchmark source only if every model answers zero probes correctly, forcing downstream systems to reason over the documents provided at evaluation time.

For each retained source, we use a Large Reasoning Model to synthesize adversarial miscitations grounded in our taxonomy (Appendix~\ref{app:verification}). The model is instructed to deeply read the paper and, for each of the five categories, generate both \emph{surface-level} miscitations falsifiable from a local passage and \emph{deep-semantic} ones requiring integration of global context, limitations, and nuanced caveats. Each candidate includes an erroneous citing sentence, a gold supporting span, a natural-language explanation, and a corrected version. We then perform independent cross-validation with a second LLM and domain experts: instances with category disagreement, ambiguous explanations, or factual inconsistencies are revised or discarded.

The resulting \mb\ benchmark has a regular structure of 254 source papers $\times$ 5 taxonomy categories $\times$ 5 instances per category, yielding 6{,}350 expert-validated miscitations with a controlled mix of surface and deep errors. It is explicitly constructed to be knowledge-blank with respect to frontier LLMs, cross-disciplinary in coverage, and aligned with the error-code space used by \ba. In Section~\ref{sec:results} we use \mb\ to evaluate both full-text baselines and \ba\ under realistic access regimes, including when cited sources are behind paywalls.

\section{\textsc{\ba}: The Bibliographic Miscitation Detection Agent}
\label{sec:method}

\ba\ is an end-to-end agentic framework that restores traceable, citation-level accountability. Instead of feeding whole paper pairs to a single long-context model, \ba\ decomposes the task into four modules: a Document Parser \& Citation Mapper (DPCM), a Cited Source Accessibility Classifier (CSAC), an Accessible Cited Source Verifier (ACSV), and an Inaccessible Cited Source Verifier (ICSV). Given a citing paper, these modules map every in-text citation to a bibliographic entry, route each cited source to the accessible or paywalled branch, and return a miscitation verdict and taxonomy-aligned error code with explicit evidence and a confidence score.

\subsection{Citation Parsing and Accessibility Routing}

The first stage of \ba\ turns heterogeneous paper formats into a citation-centric representation. A Document Parser and Citation Mapper (DPCM) parses \LaTeX, XML/HTML, and PDF into a unified hierarchical Markdown that preserves discourse structure, math, and in-text citation anchors, then builds a sentence-level citation graph linking each citing context to normalized references. Markup inputs are handled by stripping purely presentational macros, converting sectioning and math into Markdown, and rendering citation commands into their surface form while retaining anchors for the underlying bibliography keys; PDF inputs are rendered to page images and transcribed by a layout-aware multimodal model that reconstructs multi-column reading order and places equations, figures, and captions in the right locations. A lightweight Extraction Verifier audits the resulting document for heading progression, equation and float numbering, and citation index sequences, and triggers localized re-parsing when anomalies indicate OCR or layout errors, so that downstream reasoning always sees a structurally consistent paper.

On top of this representation, the Citation Mapper identifies in-text citation spans—numeric, author–year, grouped, and footnote-based—and links each to its sentence. In markup-derived documents, internal anchors yield exact bibliography mappings; in PDF-only ones, pattern-based matching is supplemented by a constrained LLM disambiguator for the small fraction of ambiguous cases. The Cited Source Accessibility Classifier (CSAC) then attempts DOI resolution, publisher full-text retrieval, and open-access surrogate matching for each referenced source. Sources with verified full text are routed to ACSV; those with only metadata go to ICSV with their metadata snapshot; and those for which neither DOI nor repository search returns a convincing match are labeled Ghost Citations and treated as Attribution \& Traceability errors. Implementation details for parsing, citation mapping, and CSAC rules are in Appendix~\ref{app:dpcm}.

\subsection{ACSV: Adaptive Multi-Stage Verification}

ACSV handles citations whose sources are fully accessible. The objective is to localize decisive evidence inside $D_{\text{cited}}$, decide whether the citing sentence $S_{\text{cite}}$ is supported, refuted, or undecidable, and do so with as little long-context reasoning as possible. Instead of sending whole documents to a single LLM call, ACSV acts as a computational funnel: inexpensive dense retrieval and NLI settle easy cases, and only genuinely ambiguous citations are escalated to a large reasoning model (LRM) for deep analysis. This design simplifies attribution of errors to specific evidence spans and substantially reduces token usage; Section~\ref{sec:results} quantifies the cost--accuracy trade-off.

Phases I--II (coarse retrieval and focused re-ranking). Let $S_{\text{cite}}$ denote the citing sentence and $D_{\text{cited}}$ the cited document. We segment $D_{\text{cited}}$ into paragraphs $\mathcal{P} = \{p_i\}_{i=1}^M$ and embed $S_{\text{cite}}$ and each $p_i$ into a shared vector space with a bi-encoder. We retrieve top-$K$ candidate paragraphs using cosine similarity
\begin{equation}
\small
    \text{Score}_{\text{retrieval}}(S_{\text{cite}}, p_i)
    = \cos(\mathbf{v}_{S_{\text{cite}}}, \mathbf{v}_{p_i})
    = \frac{\mathbf{v}_{S_{\text{cite}}} \cdot \mathbf{v}_{p_i}}{\|\mathbf{v}_{S_{\text{cite}}}\| \, \|\mathbf{v}_{p_i}\|}.
\end{equation}
A cross-encoder then re-ranks these $K$ paragraphs and selects a small focus set $\mathcal{P}_{\text{focus}}$ of top-$N$ segments. To preserve inter-sentence dependencies while avoiding paragraph-level noise, we slide a fixed-size window over each $p \in \mathcal{P}_{\text{focus}}$, producing cited-side windows $\mathcal{W}_{\text{cited}}$ of $W_{\text{size}}$ consecutive sentences with stride 1. In experiments we use $K{=}10$, $N{=}3$, and $W_{\text{size}}{=}3$, which empirically suffice to recover decisive context on \mb{} without exceeding downstream context limits.

Phase III (NLI-based logic filtering with dynamic expansion). For each window $w \in \mathcal{W}_{\text{cited}}$, we apply a Natural Language Inference (NLI) model with $w$ as \emph{premise} and $S_{\text{cite}}$ as \emph{hypothesis}:
\begin{equation}
\small
P(E, N, C \mid w, S_{\text{cite}}) = \text{NLI}(\text{premise}=w,\, \text{hypothesis}=S_{\text{cite}}),
\end{equation}
where $E$, $N$, and $C$ denote Entailment, Neutral, and Contradiction. We extract the strongest entailment and contradiction signals
\begin{align}
M_E &= \max_{w \in \mathcal{W}_{\text{cited}}} P(E \mid w, S_{\text{cite}}),\\
M_C &= \max_{w \in \mathcal{W}_{\text{cited}}} P(C \mid w, S_{\text{cite}}),
\end{align}
and apply a high-confidence early-exit rule
\begin{equation}
\label{eq:early-exit}
\text{Decision} =
\begin{cases}
\text{Correct}, & M_E > \tau_{\text{high}},\\
\text{Miscitation}, & M_C > \tau_{\text{high}}.
\end{cases}
\end{equation}
When both signals are weak or conflicting, we expand the hypothesis to include its immediate neighbors in the citing document,
\begin{equation}
    S_{\text{expanded}} = \text{concat}(S_{\text{prev}},\, S_{\text{cite}},\, S_{\text{next}}),
\end{equation}
and re-run NLI with $S_{\text{expanded}}$ as hypothesis and the same window set $\mathcal{W}_{\text{cited}}$ as premises. In our implementation we use an off-the-shelf DeBERTa-v3-large NLI model with $\tau_{\text{high}} = 0.9$ and do not fine-tune it on miscitation labels.

Phase IV (LRM deep reasoning via self-consistency). Citations that remain ambiguous after NLI enter a final, high-capacity reasoning stage. We construct a prompt containing $S_{\text{expanded}}$ and the focus paragraphs $\mathcal{P}_{\text{focus}}$, and query an LRM (e.g., \texttt{gemini-2.5-pro}~\cite{google2025gemini25card}) to decide whether the citation is supported, miscitated, or undecidable. To reduce stochastic artifacts, we employ self-consistency: we sample $M$ independent chains of thought at temperature $T$ and take the majority verdict as the final label. The associated confidence score is
\begin{equation}
    \text{Confidence}_{\text{LRM}} = \frac{\text{Count}(\text{Majority Verdict})}{M}.
\end{equation}
If this confidence falls below a safety threshold (0.6 in our experiments), ACSV outputs an Undecidable verdict and surfaces the evidence windows for manual inspection rather than forcing a brittle judgment.

\begin{figure*}[h]
  \centering
  \includegraphics[width=\linewidth]{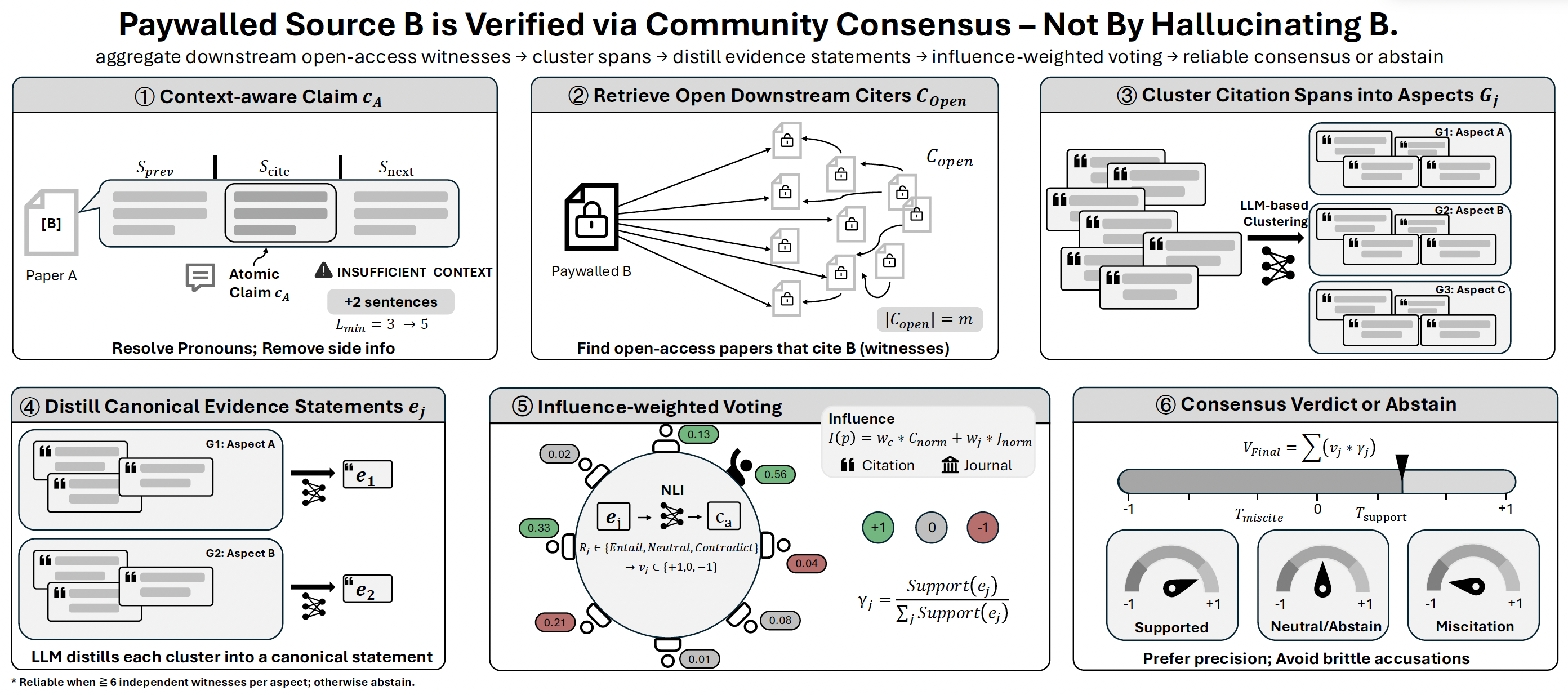}
  \caption{
  Overview of the \textbf{Inaccessible Cited Source Verifier} (ICSV) and its Evidence Committee.
  }
  \label{fig:icsv}
\end{figure*}

\subsection{ICSV: The Evidence Committee Mechanism}

ICSV handles citations whose sources are paywalled or otherwise inaccessible. Rather than attempting to reconstruct the missing full text, it treats the research community as a noisy, distributed memory of the source. The key idea is an \emph{Evidence Committee}: a set of downstream open-access papers that cite the paywalled source $B$ and whose local citation contexts are distilled into aspect-specific evidence statements. Each statement casts a weighted vote---supporting, contradicting, or remaining neutral with respect to what the citing paper $A$ claims about $B$---and ICSV aggregates these votes into a reliability-aware consensus. Algorithmic details and calibration of each stage are developed in Appendix~\ref{app:icsv} and an overview is given in Figure~\ref{fig:icsv}.

Given a citing sentence $s_A$ in paper $A$ that references $B$, ICSV first constructs a short local window $W_A$ around $s_A$ and uses an LLM to produce a single, self-contained atomic claim $c_A$ that preserves exactly what $A$ attributes to $B$ while resolving pronouns and eliding incidental context. If the model signals insufficient information, $W_A$ is gradually expanded until $c_A$ stabilizes or the case is declared underspecified. In parallel, DPCM and CSAC identify a set of downstream witness papers $C_{\text{open}}$ that cite $B$ and have accessible full text. For each explicit in-text citation to $B$ in each witness, we apply the same claim-preserving procedure, obtaining witness claims $\{c_1,\dots,c_m\}$ that collectively encode how the community describes $B$.

To organize these witness claims, we apply LLM-based semantic clustering that groups $\{c_\ell\}$ into coherent aspects $G_1,\dots,G_k$ of $B$ (for example, algorithmic contribution, dataset, or key empirical finding). Within each cluster $G_j$ an LLM distills a single canonical evidence statement $e_j$ that captures only the overlap among its claims and preserves qualifiers and scope. We then assign each witness paper $p$ a field-normalized influence score $\mathcal{I}(p)$ that combines its paper-level citation percentile and venue-level standing. Let $\mathrm{IF}(p)$ be the impact factor of $p$'s venue and $\mathrm{Cite}(p)$ its citation count. We compute
\begin{align}
    J_{\text{norm}}(p) &= \mathrm{Rank}_{\%}\big( \mathrm{IF}(p) \mid \mathrm{Field}(p) \big), \\
    C_{\text{norm}}(p) &= \mathrm{Rank}_{\%}\big( \mathrm{Cite}(p) \mid \mathrm{Field}(p), \mathrm{Year}(p) \big),
\end{align}
and aggregate them as
\begin{equation}
    \mathcal{I}(p) = w_c \cdot C_{\text{norm}}(p) + w_j \cdot J_{\text{norm}}(p),
\end{equation}
with fixed weights $w_c = 0.6$ and $w_j = 0.4$, so that paper-level influence dominates venue prestige and highly cited work in niche venues can still carry substantial weight. The credibility of an evidence statement $e_j$ is the normalized sum of influences from the unique witness papers that contribute to its cluster:
\begin{align}
    \mathrm{Support}(e_j) &= \sum_{s \in G_j} \mathcal{I}(\mathrm{SourcePaper}(s)), \\
    \gamma_j &= \frac{\mathrm{Support}(e_j)}{\sum_{i=1}^k \mathrm{Support}(e_i)}.
\end{align}

ICSV then compares each $e_j$ to the citing-side claim $c_A$. An LLM classifies the relation between $e_j$ and $c_A$ as entailment, contradiction, or neutral, which we map to a scalar vote $v_j \in \{+1,0,-1\}$. The Evidence Committee's credibility-weighted consensus score is
\vspace{-1em}
\begin{equation}
    \mathcal{V}_{\text{final}} = \sum_{j=1}^k v_j \cdot \gamma_j,
\end{equation}
\vspace{-0.8em}

with $\mathcal{V}_{\text{final}} \in [-1,1]$. We label the citation as Supported if $\mathcal{V}_{\text{final}} > T_{\text{support}}$, Miscitation if $\mathcal{V}_{\text{final}} < T_{\text{miscite}}$, and otherwise Undecidable; in all experiments we set $T_{\text{support}} = 0.3$ and $T_{\text{miscite}} = -0.3$. To prioritize precision in high-stakes settings, ICSV further enforces a reliability-aware abstention rule: if the number of distinct witnesses $|C_{\text{open}}|$ falls below a minimum committee size ($K_{\min} = 6$) or if the credibility mass is split across incompatible relations, the system explicitly abstains with a low-confidence Undecidable verdict rather than overinterpreting sparse or conflicting community evidence. All intermediate artifacts---$c_A$, the witness set, aspect clusters, evidence statements, and per-cluster votes---are logged, making paywall-regime decisions as traceable as those in the accessible regime.

\subsection{Taxonomy-Aligned Labeling}
\label{subsec:taxonomy_labeling}

Beyond binary validity, \ba\ aims to characterize \emph{how} a citation fails, using the five-category taxonomy of Section~\ref{subsec:taxonomy} as its error-code space. For every citation that ACSV or ICSV labels as a miscitation, a lightweight taxonomy classifier assigns a single primary category. The classifier operates over the same evidence bundle used for verification: the expanded citing context $S_{\text{expanded}}$, the retrieved evidence windows $\mathcal{W}_{\text{cited}}$ (for ACSV) or distilled evidence statements $\{e_j\}$ (for ICSV), and CSAC metadata such as article type and retraction status. To enforce the Dependency Precedence Rule, the classifier is prompted with a hierarchical decision tree that first checks Attribution \& Traceability, then Citation Validity, then Content Misrepresentation, and only then Scope Extrapolation and Evidence Characterization. We instantiate it with a compact LLM (\texttt{gpt-4o-2024-08-06}~\cite{hurst2024gpt}) and use a small self-consistency ensemble that returns the majority category together with a short natural-language rationale. The final output for each citation is thus a validity verdict, a taxonomy-aligned error code, an evidence bundle, and a calibrated confidence score.

\section{Experiments and Results}
\label{sec:results}

\subsection{Setup and Evaluation Regimes}

We evaluate \ba on \mb under two complementary regimes that mirror realistic deployment. In \textbf{MisciteBench-Open}, the full text of each cited source is provided and instances are routed to ACSV; the task is to decide whether the citing statement is supported or miscitated and to return a taxonomy-aligned diagnosis. In \textbf{MisciteBench-Paywall}, the cited source text is withheld and only bibliographic metadata plus downstream open-access citers are available, so instances are routed to ICSV.

Because deep semantic miscitations become intrinsically underdetermined when the source cannot be read, the paywall regime focuses on the surface-level subset of \mb (3{,}810 instances), where verification can in principle be resolved from community evidence alone. The construction of \mb, the surface/deep split, and its contamination-controlled, knowledge-blank design are detailed in Section~\ref{subsec:miscitebench} and Appendix~\ref{app:taxonomy_annotation}.

\vspace{-0.25cm}
\subsection{Baselines and Backbone-Controlled Comparisons}

Direct comparison to earlier work on citation sentiment or scientific claim verification is limited by missing code, short-context assumptions, and incompatible inputs (abstract-only, snippet-level, or domain-specific settings). To avoid over-claiming transfer, we construct \emph{backbone-controlled} baselines: every method uses exactly the same underlying LLM/LRM, but with different verification architectures.

For accessible sources we use a \textbf{Full-Text} baseline that concatenates the citing context and the entire cited paper into a single prompt and asks the backbone to decide \emph{Supported} vs.\ \emph{Miscitation} and explain its decision. For paywalled sources we use a \textbf{Search} baseline that allows the backbone to call web-search tools, retrieve potential surrogates or snippets, and then verify against whatever it finds. \ba uses the same backbones inside ACSV (Open) and ICSV (Paywall), but replaces monolithic prompting with a retrieval--NLI--reasoning funnel and the Evidence Committee mechanism. This design isolates the contribution of the agentic architecture: given the same model, can \ba convert long-context and paywall verification from a brittle prompt into a reliable procedure?

\subsection{Metrics}

Our primary metric is \textbf{Acc-pass@3}, which jointly evaluates correctness and diagnostic faithfulness. For each instance and method we generate three independent samples. A method succeeds on that instance if at least one sample both predicts the correct validity label (\emph{Supported} vs.\ \emph{Miscitation}) and provides an explanation that an independent LLM grader judges semantically equivalent to the gold miscitation rationale. All other cases, including abstentions and partially correct rationales, are counted as errors. Appendix~\ref{app:eval_details} gives the formal definition, grader prompts, and human validation of the grading pipeline.

For MisciteBench-Open we also report \textbf{Token Economy}, the relative reduction in total tokens processed (inputs and outputs) compared to the same-backbone Full-Text baseline, computed on instances where both methods output a non-abstaining verdict:
\[
\mathrm{TokenEcon} = 1 - \frac{\mathrm{Tok}_\ba}{\mathrm{Tok}_{\text{FT}}}.
\]
Details on token accounting and decoding configurations are provided in Appendix~\ref{app:eval_details}.

\subsection{Accessible Sources: MisciteBench-Open}

Table~\ref{tab:open_results} reports results when the full texts of cited sources are accessible. Across all backbones, \ba consistently improves Acc-pass@3 over the corresponding Full-Text baseline while substantially reducing computation. Gains in diagnostic accuracy range from \textbf{+5.7} to \textbf{+19.8} absolute points, with the largest improvements for models that are known to be fragile under long context (e.g., \texttt{gpt-4o} and mid-sized open-weight backbones). At the same time, \ba reduces token usage by \textbf{44.6--79.4\%}, confirming that the retrieval--NLI funnel resolves most citations without invoking expensive long-context reasoning.

Qualitatively, Full-Text failures concentrate on citations whose decisive evidence is split across several neighboring sentences with caveats or limitations, or entangled with other contributions inside dense paragraphs. In these cases the backbone often produces fluent but weakly grounded rationales: the correct label may be guessed, but the explanation drifts away from the actual evidence. By forcing verification to proceed through a high-recall evidence pool, local NLI filtering, and selective escalation, ACSV anchors decisions to a small, explicit set of windows and reduces both long-context dilution and explanation drift.

\begin{table}[!t]
\caption{MisciteBench-Open: miscitation detection and diagnosis when full texts of cited sources are accessible. Token Econ reports the percentage of tokens saved by \ba relative to the corresponding Full-Text baseline.}
\label{tab:open_results}
\centering
\footnotesize
\setlength{\tabcolsep}{3pt}
\renewcommand{\arraystretch}{0.95}

\sisetup{
  detect-all,
  table-number-alignment = center
}

\newcommand{\HdrShift}{-0.65ex}
\newcommand{\Hdr}[1]{\raisebox{\HdrShift}{\strut\textbf{#1}}}

\resizebox{\linewidth}{!}{%
\begin{tabular}{l c S[table-format=2.1] S[table-format=3.1] S[table-format=2.1]}
\toprule
\multirow{2}{*}{\Hdr{Model}} &
\multirow{2}{*}{\Hdr{Org.}} &
\multicolumn{2}{c}{\textbf{Acc-pass@3 $\uparrow$}} &
\multicolumn{1}{c}{\multirow{2}{*}{\Hdr{Token Econ $\uparrow$}}} \\
\cmidrule(lr){3-4}
& & \multicolumn{1}{c}{\textbf{Full-Text}} & \multicolumn{1}{c}{\textbf{\ba}} & \\
\midrule

gpt-5-2025-08-07     & \multirow{5}{*}{\OpenAIIcon} & 92.1 & {\bfseries 98.8} & 65.3 \\
o4-mini-2025-04-16   &                             & 88.4 & {\bfseries 96.4} & 79.4 \\
gpt-4o-2024-08-06    &                             & 84.6 & {\bfseries 94.8} & 60.2 \\
gpt-oss-120b         &                             & 82.9 & {\bfseries 92.7} & 59.3 \\
gpt-oss-20b          &                             & 72.4 & {\bfseries 87.6} & 70.0 \\
\midrule

claude-sonnet-4-20250514 & \multirow{2}{*}{\AnthropicIcon} & 91.3 & {\bfseries 97.8} & 74.7 \\
claude-opus-4-20250514   &                                & 94.3 & {\bfseries 100.0} & 44.6 \\
\midrule

gemini-2.5-pro   & \multirow{3}{*}{\GeminiIcon} & 95.8 & {\bfseries 97.9} & 74.8 \\
gemini-3-flash   &                             & 93.1 & {\bfseries 96.4} & 66.2 \\
gemini-3-pro     &                             & 96.8 & {\bfseries 100.0} & 64.7 \\
\midrule

Nemotron 3 Nano (30B A3B)              & \multirow{2}{*}{\NVIDIAIcon} & 73.0 & {\bfseries 88.4} & 72.1 \\
Llama-3.3 Nemotron Super 49B v1        &                             & 80.8 & {\bfseries 89.9} & 66.4 \\
\midrule

Qwen3-235B-A22B-Thinking-2507 & \multirow{3}{*}{\QwenIcon} & 86.4 & {\bfseries 95.5} & 61.8 \\
Qwen3 VL 32B Thinking         &                           & 79.3 & {\bfseries 92.4} & 66.6 \\
Qwen3 VL 8B Thinking          &                           & 54.8 & {\bfseries 80.2} & 76.5 \\
\midrule

Ministral 3 (14B Reasoning 2512) & \multirow{2}{*}{\MistralIcon} & 56.2 & {\bfseries 82.1} & 74.5 \\
Magistral Medium                  &                              & 66.4 & {\bfseries 87.0} & 68.3 \\
\midrule

Deepseek-V3.2 (Thinking)      & \multirow{4}{*}{\DeepSeekIcon} & 88.0 & {\bfseries 98.6} & 62.4 \\
DeepSeek-R1-0528              &                                & 92.4 & {\bfseries 97.2} & 62.1 \\
Deepseek R1 Distill Qwen 32B  &                                & 76.6 & {\bfseries 92.0} & 68.7 \\
Deepseek R1 Distill Qwen 14B  &                                & 68.4 & {\bfseries 88.1} & 72.9 \\
\midrule

Llama 3.1 405B Instruct & \multirow{2}{*}{\MetaIcon} & 74.5 & {\bfseries 91.0} & 60.2 \\
Llama 3.3 70B Instruct  &                           & 70.3 & {\bfseries 84.6} & 63.1 \\
\bottomrule
\end{tabular}%
}

\end{table}

\subsection{Paywalled Sources: MisciteBench-Paywall}

Table~\ref{tab:paywall_results} summarizes results when the cited source is inaccessible. The Search baseline, despite being allowed to call web search and retrieve open-access surrogates, achieves only \textbf{22.1--36.2} Acc-pass@3 on the strongest backbones and substantially less on smaller models. In practice, retrieved snippets are often incomplete, version-mismatched, or only loosely related to the actual citation, and explanations frequently over-interpret these fragments to fit the query.

In contrast, ICSV more than doubles performance across all backbones while operating entirely offline, without any live search. Acc-pass@3 rises to \textbf{66.5--80.3} on the same backbones, and similar relative gains hold for smaller models. Rather than hallucinating the contents of paywalled sources, \ba\  reconstructs what the community collectively attributes to each source via its downstream citers, aggregates this evidence with field-normalized credibility weights, and abstains explicitly when community memory is too sparse or inconsistent. This supports the central design choice of ICSV: paywalled miscitation detection should be framed as a reliability problem over community evidence, not as a brittle retrieval problem.

\begin{table}[!t]
\caption{MisciteBench-Paywall: miscitation detection when the source text is inaccessible. Acc-pass@3 is reported for a retrieval-augmented baseline (Search) and \ba. Models marked with $^{\ast}$ use a self-implemented retrieval toolchain for Search, while unmarked models use an API-native toolchain. \ba uses only ICSV's Evidence Committee, without any external search.}
\label{tab:paywall_results}
\centering
\footnotesize
\setlength{\tabcolsep}{3pt}
\renewcommand{\arraystretch}{0.6}

\sisetup{
  detect-all,
  table-number-alignment = center
}

\newcommand{\HdrShift}{-0.65ex}
\newcommand{\Hdr}[1]{\raisebox{\HdrShift}{\strut\textbf{#1}}}

\resizebox{\linewidth}{!}{%
\begin{tabular}{l c S[table-format=2.1] S[table-format=2.1]}
\toprule
\multirow{2}{*}{\Hdr{Model}} &
\multirow{2}{*}{\Hdr{Org.}} &
\multicolumn{2}{c}{\textbf{Acc-pass@3 $\uparrow$}} \\
\cmidrule(lr){3-4}
& & \multicolumn{1}{c}{\textbf{Search}} & \multicolumn{1}{c}{\textbf{\ba}} \\
\midrule

gpt-5-2025-08-07     & \multirow{5}{*}{\OpenAIIcon} & 36.2 & {\bfseries 80.3} \\
o4-mini-2025-04-16   &                             & 29.8 & {\bfseries 66.4} \\
gpt-4o-2024-08-06    &                             & 22.1 & {\bfseries 66.5} \\
gpt-oss-120b$^{\ast}$&                             & 16.4 & {\bfseries 69.4} \\
gpt-oss-20b$^{\ast}$ &                             &  6.6 & {\bfseries 55.7} \\
\midrule

claude-sonnet-4-20250514 & \multirow{2}{*}{\AnthropicIcon} & 30.2 & {\bfseries 66.8} \\
claude-opus-4-20250514   &                                & 34.4 & {\bfseries 74.2} \\
\midrule

Nemotron 3 Nano (30B A3B)$^{\ast}$       & \multirow{2}{*}{\NVIDIAIcon} & 12.3 & {\bfseries 56.2} \\
Llama-3.3 Nemotron Super 49B v1$^{\ast}$ &                             & 11.6 & {\bfseries 59.3} \\
\midrule

Qwen3-235B-A22B-Thinking-2507$^{\ast}$ & \multirow{3}{*}{\QwenIcon} & 18.2 & {\bfseries 69.2} \\
Qwen3 VL 32B Thinking$^{\ast}$         &                           & 13.7 & {\bfseries 63.9} \\
Qwen3 VL 8B Thinking$^{\ast}$          &                           &  2.1 & {\bfseries 54.2} \\
\midrule

Ministral 3 (14B Reasoning 2512) & \multirow{2}{*}{\MistralIcon} & 17.9 & {\bfseries 56.3} \\
Magistral Medium                 &                              & 28.0 & {\bfseries 62.4} \\
\midrule

Deepseek-V3.2 (Thinking)$^{\ast}$     & \multirow{4}{*}{\DeepSeekIcon} & 14.8 & {\bfseries 76.8} \\
DeepSeek-R1-0528$^{\ast}$             &                                & 14.5 & {\bfseries 70.6} \\
Deepseek R1 Distill Qwen 32B$^{\ast}$ &                                &  9.6 & {\bfseries 62.7} \\
Deepseek R1 Distill Qwen 14B$^{\ast}$ &                                &  8.2 & {\bfseries 59.6} \\
\midrule

Llama 3.1 405B Instruct$^{\ast}$ & \multirow{2}{*}{\MetaIcon} & 10.3 & {\bfseries 64.3} \\
Llama 3.3 70B Instruct$^{\ast}$  &                           &  7.1 & {\bfseries 59.4} \\
\bottomrule
\end{tabular}%
}

\vspace{0.1cm}

\end{table}

\subsection{Evidence Committee Reliability}
\label{sec:committee_reliability}

\begin{figure}[t]
    \centering
    \includegraphics[width=0.9\linewidth]{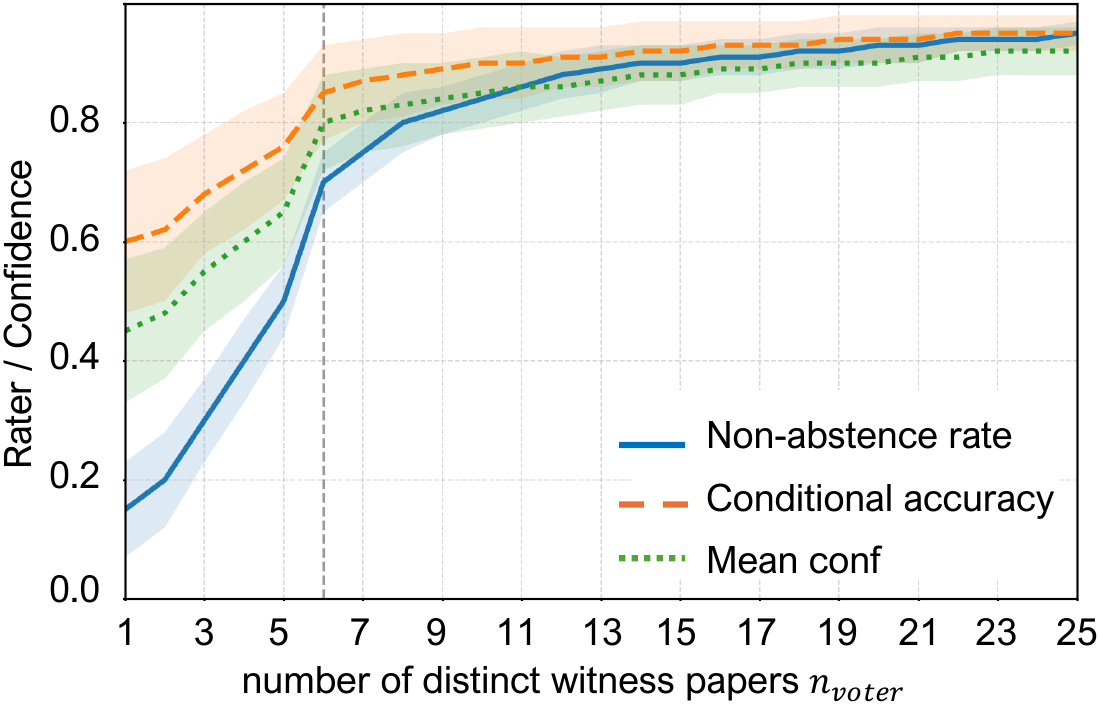}
    \caption{Evidence Committee behavior as a function of the number of distinct downstream witness papers whose claims fall into the dominant aspect cluster $G_{j^\star}$ (and thus support its canonical evidence statement $e_{j^\star}$) for a paywalled source. Once six or more such witnesses are available, ICSV's decisions become both frequent and highly accurate; below this regime it abstains more.}
    \label{fig:icsv_voters_ablation}
\end{figure}

To understand when community evidence is strong enough to support decisive paywall judgments, we ablate ICSV as a function of the number of distinct downstream witness papers that contribute to the \emph{dominant} aspect cluster $G_{j^\star}$ and its canonical evidence statement $e_{j^\star}$ for each paywalled citation. For this analysis we restrict to paywalled citations where the evidence statement that is directly comparable to the citing claim admits an evidence committee of at least 25 distinct downstream witnesses, yielding a high-signal pool of 62 paywalled sources whose committees we systematically subsample. Figure~\ref{fig:icsv_voters_ablation} plots the non-abstention rate, conditional accuracy, and calibrated confidence against this per-aspect committee size.

Across backbones and disciplines we observe a sharp reliability transition. When the dominant aspect cluster of a paywalled source is supported by only one or two witness papers contributing to $e_{j^\star}$, ICSV correctly treats most cases as underspecified and frequently returns \emph{Undecidable}. With three to five such witnesses, both accuracy and confidence improve but remain more variable. Once the dominant aspect has at least six independent witnesses in its cluster, non-abstention rates and conditional precision stabilize at high levels and additional witnesses yield only marginal gains. This empirical knee motivates the global committee-size threshold $K_{\min}=6$ in ICSV's reliability-aware abstention rule and aligns with the intuition that robust paywall verification requires multiple, partly redundant community views. Appendix~\ref{app:committee_reliability} provides the full ablation protocol.

\section{Conclusion}
\label{sec:conclusion}

This work advances bibliographic integrity by making miscitation a first-class, auditable object rather than an anecdotal pathology. We introduce a five-category taxonomy that turns vague complaints about ``bad citations'' into a precise error-code space, instantiate it at scale in \mb, and operationalize it in \ba, an agentic verifier that unifies accessible and paywalled regimes. On accessible sources, \ba's adaptive zoom-in architecture matches or surpasses full-document LLM prompting while cutting token usage by as much as \textbf{79.4\%}; on inaccessible sources, its Evidence Committee more than doubles miscitation-detection accuracy over search-augmented baselines without ever fabricating the contents of paywalled articles. Together, these components demonstrate that citation chains can be checked systematically, with traceable evidence and taxonomy-aligned diagnoses, rather than sampled opportunistically.

More broadly, \ba points toward self-auditing scientific workflows in which drafting and verification are decoupled, paywalls are treated as an evidential setting rather than a hard failure, and citation integrity becomes measurable, diagnosable, and ultimately controllable at scale. Extended discussion of integration into authoring tools and editorial pipelines, applicability beyond scholarly articles, and remaining limitations appears in Appendix~\ref{app:extended_discussion}.

\newpage
\clearpage

\section*{Impact Statement}

Citations are not merely stylistic conventions; they are the operational substrate of scientific authority. When a citation is wrong—because the source is fabricated, misquoted, overgeneralized, or mischaracterized—the damage is not local to a paragraph: it propagates through reviews, meta-analyses, clinical guidelines, and downstream models trained on the literature. This risk has become acute in the era of generative AI, where fluent writing can be paired with plausible-but-hallucinated or subtly distorted references at scale. At the same time, systematic verification has remained structurally infeasible: human review does not scale with modern citation volumes, and automated tools have historically collapsed under long full texts, narrow domains, or the paywall barrier that hides precisely the sources most in need of scrutiny.

This paper targets that bottleneck directly by making miscitation a first-class, \emph{traceable} object. \textsc{BibAgent} (\ba) operationalizes citation verification as a two-regime evidential problem—\emph{accessible} versus \emph{inaccessible} sources—rather than treating paywalls as a hard failure mode. In the accessible regime, \ba's adaptive retrieval--NLI--reasoning funnel localizes decisive evidence windows and escalates to high-capacity reasoning only when needed, improving diagnostic accuracy while substantially reducing computation. In the paywalled regime, \ba's \emph{Evidence Committee} reconstructs what the research community collectively attributes to an inaccessible source via downstream citation contexts, aggregates these with reliability-aware weighting, and abstains when evidence is sparse or inconsistent. Together with \textsc{MisciteBench} (\mb)—a large, contamination-controlled, cross-disciplinary benchmark aligned with a unified five-category taxonomy—this work enables systematic evaluation and, more importantly, makes citation integrity \emph{measurable, diagnosable, and actionable} across fields.

The positive scientific impact is immediate and structural. For authors, \ba can function as a pre-submission ``citation firewall'' that detects misalignments before they enter the scholarly record, reducing inadvertent errors and discouraging strategic citation inflation. For reviewers and editors, it converts citation checking from an intractable obligation into a targeted process: evidence is surfaced, error types are coded, and uncertainty is explicit, allowing scarce expert attention to be spent where it matters most. For the broader ecosystem, \ba supports scalable audits of citation quality across venues and disciplines, enabling meta-research on how misinformation, retractions, and overstated claims spread through citation networks. In high-stakes domains—e.g., biomedical or policy-facing literature—systematic miscitation detection can reduce the probability that incorrect evidence chains influence real-world decisions. Finally, by treating paywalls as an evidential setting rather than an excuse for abstention, this work offers a concrete mechanism to narrow a long-standing integrity blind spot in scientific publishing.

There are also meaningful risks. A miscitation detector could be misused as an automated weapon for policing, harassment, or reputational attacks, especially if outputs are treated as definitive rather than probabilistic. False positives may disproportionately affect certain disciplines, writing styles, or non-native authors; and the paywall regime's community-consensus reconstruction can, in principle, reproduce collective misconceptions. Moreover, the influence-weighted aggregation in the Evidence Committee, if deployed naively, could amplify prestige bias by over-valuing highly cited venues or established communities.

We mitigate these risks through design choices that prioritize \emph{auditable restraint} over forced certainty: \ba exposes concrete evidence windows or committee statements, logs intermediate artifacts, provides calibrated confidence, and explicitly abstains under low committee size or inconsistent signals—reducing incentives to over-interpret weak evidence. The taxonomy-aligned single primary error code (via dependency precedence) is intended to support reproducible checking workflows, not punitive automation. We recommend deployments that keep a human in the loop for consequential decisions, prohibit use as a sole basis for sanctions, and report uncertainty alongside verdicts. Future work should additionally audit performance and abstention behavior across disciplines and writing demographics, stress-test consensus failure modes, and explore alternative weighting schemes that reduce prestige amplification while preserving reliability. In sum, this paper advances the infrastructure for scientific self-correction by making citation integrity scalable, transparent, and robust to the realities of paywalled scholarship—while explicitly acknowledging and designing against misuse.




\nocite{langley00}

\bibliography{example_paper}
\bibliographystyle{icml2026}

\newpage
\appendix
\onecolumn
\section{Taxonomy Annotation Protocol}
\label{app:taxonomy_annotation}

This appendix provides (i) the full five-category miscitation taxonomy used in the main text and (ii) the details of the expert annotation study that validates its applicability across disciplines.

\subsection{Unified 5-Category Taxonomy}

Conceptually, our taxonomy decomposes miscitation along five orthogonal dimensions---\emph{status of the source}, \emph{factual content}, \emph{scope of application}, \emph{evidence strength}, and \emph{attribution link}---which jointly cover all failure modes observed in our corpus. Table~\ref{tab:miscite_taxonomy_mapping} summarizes how legacy labels from prior miscitation studies are absorbed into these categories.

\begin{table}[H]
\caption{Mapping from legacy miscitation labels to our unified 5-category taxonomy.}
\label{tab:miscite_taxonomy_mapping}
\centering
\small
\setlength{\tabcolsep}{4pt}
\begin{tabularx}{\linewidth}{
  >{\raggedright\arraybackslash}p{0.22\linewidth} |
  >{\raggedright\arraybackslash}X
  >{\raggedright\arraybackslash}X
}
\toprule
\textbf{New Category} &
\textbf{Core Litmus Question} &
\textbf{Absorbed Legacy Subtypes} \\
\midrule
Citation Validity Error &
\emph{Is the source itself disqualified from serving as scientific evidence?} &
Obsolete or Retracted Citation; Secondary-Source Misuse \\
\\
Content Misrepresentation Error &
\emph{Does the source, when read in context, actually say what the citing text claims it says?} &
Irrelevant Citation; Contradictory Citation; Selective Quotation; Omitted Qualifiers; Conflated Findings \\
\\
Scope Extrapolation Error &
\emph{Is an otherwise valid conclusion being applied outside the population, setting, or task for which it was established?} &
Overgeneralization; Methodological Misapplication; Data-Scope Misuse \\
\\
Evidence Characterization Error &
\emph{Is the type/strength of evidence claimed (e.g., causal, definitive) actually supported by the study design and statistics?} &
Correlation-as-Causation; Statistical or Metrical Distortion \\
\\
Attribution \& Traceability Error &
\emph{Can a reader reliably locate and correctly attribute the source using the provided citation metadata?} &
Ghost Citation; Author Misattribution \\
\bottomrule
\end{tabularx}
\vspace{0.2cm}

\end{table}

Starting from a survey of existing typologies and a manual audit of hundreds of real-world miscitations sampled across all 21 high-level disciplines, we consolidate previously scattered subtypes into five conceptually distinct categories. Each category is defined by a diagnostic litmus question that makes annotation operational and reproducible:

\begin{enumerate}
    \item \textbf{Citation Validity Error (Status of the Source).}
    The cited work itself lacks qualification as scientific evidence---for example, it has been retracted, superseded, or is a secondary source (e.g., a review or meta-analysis) being cited as if it were primary experimental evidence. The error concerns the \emph{status} of the source, not its content. Litmus question: \emph{\textbf{``Is the source itself disqualified from serving as scientific evidence?''}} This category subsumes \emph{Obsolete or Retracted Citation} and \emph{Secondary-Source Misuse}.

    \item \textbf{Content Misrepresentation Error (Factual Content).}
    The citing text substantively distorts, fabricates, or reverses the findings, arguments, or conclusions of the source. Examples include citing a topically unrelated paper as evidence (\emph{Irrelevant Citation}), citing a refutation as support (\emph{Contradictory Citation}), or selectively quoting while omitting key qualifiers so that the meaning changes (\emph{Selective Quotation}, \emph{Omitted Qualifiers}, \emph{Conflated Findings}). Litmus question: \emph{\textbf{``If I read the source in context, does it actually say what the citing sentence claims it says?''}}

    \item \textbf{Scope Extrapolation Error (Scope of Application).}
    The source is correctly understood, but its conclusion is applied beyond the populations, settings, tasks, or methods for which it was established. Typical cases include \emph{Overgeneralization} from narrow samples to broad populations, \emph{Methodological Misapplication} outside validated constraints, and \emph{Data-Scope Misuse} that promotes a subset analysis to a claim about the full dataset. Litmus question: \emph{\textbf{``Is an otherwise valid conclusion being applied outside the population, setting, or task for which it was established?''}}

    \item \textbf{Evidence Characterization Error (Evidence Strength).}
    The citing text mischaracterizes the logical type, strength, or certainty of the evidence in the source. This includes treating correlational findings as causal (\emph{Correlation-as-Causation}) or exaggerating statistical evidence (\emph{Statistical or Metrical Distortion}), such as describing marginal effects as ``conclusive causal evidence.'' Litmus question: \emph{\textbf{``Is the type and strength of evidence claimed in the citing text actually supported by the source’s study design and statistics?''}}

    \item \textbf{Attribution \& Traceability Error (Attribution Link).}
    Errors in citation metadata break the link between claim and source. Examples include non-existent or unresolvable references (\emph{Ghost Citation}) and assigning a result to the wrong author or paper (\emph{Author Misattribution}). The error concerns the citation as a scholarly signpost, not the underlying evidence. Litmus question: \emph{\textbf{``Could a reader, using only this citation metadata, reliably locate the correct source and author of the claimed idea?''}}
\end{enumerate}

These five categories were sufficient to label all 6{,}350 miscitation instances in \mb\ without an ``other'' bucket. Annotators assign exactly one primary category per instance, guided by the litmus questions above. When multiple error types co-occur, we enforce a logical \emph{Dependency Precedence Rule} that mirrors the verification process: checking a citation aborts at the first point of failure. The precedence order is
\begin{center}
    \small
    \emph{Attribution \& Traceability} $\rightarrow$ \emph{Citation Validity} $\rightarrow$ \emph{Content Misrepresentation} $\rightarrow$ \emph{Scope Extrapolation} $\rightarrow$ \emph{Evidence Characterization}.
\end{center}
For example, if the citation metadata is unusable (\emph{Attribution}), one cannot assess retraction status (\emph{Validity}) or content fidelity (\emph{Content}), so Attribution becomes the primary label.

\subsection{Sampling Strategy and Annotator Panel}

The goals of the expert annotation study are twofold: (i) to test whether the five-category taxonomy is empirically usable and complete across disciplines, and (ii) to quantify how reliably independent experts can reproduce each other's judgements when constrained by the Dependency Precedence Rule.

\paragraph{Instance selection.}
From the full \mb{} benchmark of 6{,}350 miscitation instances, we draw a stratified sample of $N = 500$ cases for manual validation. Stratification is performed jointly over:
\begin{enumerate}
    \item \textbf{Taxonomy category}: we maintain approximately equal representation of the five error types (Citation Validity, Content Misrepresentation, Scope Extrapolation, Evidence Characterization, Attribution \& Traceability).
    \item \textbf{Discipline}: we preserve the distribution over the 21 high-level disciplines used in \mb{} construction (e.g., Agricultural Sciences, Clinical Medicine, Computer Science, Social Sciences, Visual \& Performing Arts), so that no single field dominates the sample.
\end{enumerate}
This design ensures that the annotation study probes both the breadth of scientific domains and the full error-code space, rather than concentrating on a narrow subset of easy or homogeneous instances.

\paragraph{Annotator qualifications.}
Each instance is independently labeled by $3$ experts with doctoral-level or advanced graduate training in a subfield relevant to the source paper. We restrict assignment so that:
\begin{itemize}
    \item every annotator has prior experience with reading and evaluating peer-reviewed scientific articles in the corresponding discipline, and
    \item no annotator is asked to label instances from a field outside their demonstrated area of expertise.
\end{itemize}
Experts are blind to the synthetic origin of \mb{} and are instructed to treat each instance exactly as they would when reviewing a real manuscript.

\subsection{Annotation Materials and User Interface}

For each candidate miscitation instance, annotators are provided with a compact but fully contextualized bundle:

\begin{itemize}
    \item \textbf{Source-side context}: the title, abstract, and gold supporting excerpt (i.e., the span in the source paper that the miscitation was constructed to distort).
    \item \textbf{Citing-side context}: the expanded citing window $S_{\text{expanded}}$ (the citing sentence plus its immediate neighbors), matching the context used by \ba{} during verification.
    \item \textbf{Taxonomy reference}: the definitions of the five categories, including their diagnostic litmus questions and short examples, identical to those in Section~\ref{subsec:taxonomy}.
\end{itemize}

The annotation interface is a structured form that guides experts through the same logical decision path used by \ba{}. For each instance, the interface:
\begin{enumerate}
    \item presents the citing context and source excerpt side-by-side,
    \item displays the five taxonomy categories with their litmus questions in a collapsible panel, and
    \item enforces the Dependency Precedence Rule via a hierarchical selection widget (described below).
\end{enumerate}

\subsection{Decision Procedure and Dependency Precedence}

Annotators are explicitly instructed to mimic the stepwise verification process a careful reviewer would perform. The interface enforces the following sequence:

\begin{enumerate}
    \item \textbf{Attribution \& Traceability check.} Determine whether, given the citation metadata, a reader could reliably locate and identify the correct source (Attribution \& Traceability Error). If an Attribution failure is present (e.g., ghost citation, hopelessly ambiguous metadata), annotators select this label and no further categories can be chosen.
    \item \textbf{Citation Validity check.} If the citation is traceable, decide whether the source itself remains valid as scientific evidence (Citation Validity Error), e.g., retracted or misused secondary source. If so, this label is selected and lower levels are disabled.
    \item \textbf{Content Misrepresentation check.} If the source is valid, compare the citing context to the source excerpt and decide whether the citing text faithfully represents the factual content (Content Misrepresentation Error).
    \item \textbf{Scope Extrapolation check.} If content is correctly represented, decide whether the citing text applies an otherwise valid conclusion outside the population, setting, or task for which it was established (Scope Extrapolation Error).
    \item \textbf{Evidence Characterization check.} Finally, decide whether the citing text mischaracterizes the logical type or strength of the evidence (Evidence Characterization Error), e.g., treating correlational results as causal or overstating statistical certainty.
\end{enumerate}

This procedure operationalizes the \emph{Dependency Precedence Rule} described in Section~\ref{subsec:taxonomy}: once an annotator records a failure at a higher level (e.g., Attribution), all lower levels (Validity, Content, Scope, Evidence) are automatically grayed out and cannot be selected for that instance. Conversely, if no failure is detected at any level, the annotator records the instance as \emph{No miscitation}.

In addition to the five primary taxonomy labels, the interface provides two auxiliary options:
\begin{itemize}
    \item \textbf{Other}: the annotator judges that there is a genuine miscitation, but it does not fit any of the five categories.
    \item \textbf{Uncertain}: the annotator cannot confidently decide due to ambiguity or insufficient information.
\end{itemize}
For each chosen label (including ``Other'' and ``Uncertain''), annotators supply a short free-text rationale explaining their decision. These rationales are later used during adjudication and for qualitative error analysis.

\subsection{Agreement Measurement}

To quantify reproducibility, we compute pairwise Cohen's $\kappa$ among the three annotators over the five taxonomy labels only, ignoring instances labeled as ``Other'' or ``Uncertain'' by any annotator.\footnote{We exclude ``Other'' and ``Uncertain'' from the primary agreement analysis because they are by design fallback categories; their frequency is reported separately.} For annotators $a$ and $b$, Cohen's $\kappa$ is defined as:
\begin{equation}
    \kappa = \frac{p_o - p_e}{1 - p_e},
\end{equation}
where $p_o$ is the observed label agreement and $p_e$ is the expected agreement under chance, computed from the empirical label marginals.

Across annotator pairs, we obtain an average $\kappa = 0.73$, which falls in the ``substantial'' agreement range under conventional benchmarks. This indicates that, given the same source and citing contexts and a shared taxonomy, independent experts largely converge on the same primary error category.

The marginal usage rates of the fallback options are low: ``Other'' is selected in \textbf{$4.2\%$} of annotations and ``Uncertain'' in \textbf{$3.6\%$}. These small rates, together with the substantial $\kappa$, provide empirical evidence that:
\begin{enumerate}
    \item the five-category taxonomy is sufficiently expressive for the \mb{} corpus, and
    \item the operational litmus questions make category boundaries clear enough for reproducible annotation.
\end{enumerate}

\subsection{Adjudication and Final Labels}

After individual annotation, we perform a structured adjudication step to obtain a single gold label per instance:

\begin{itemize}
    \item \textbf{Simple majority.} If at least two annotators agree on the same taxonomy category, that category is taken as the final label.
    \item \textbf{Complete disagreement or fallback use.} If all three annotators select different categories, or if any annotator chooses ``Other'' or ``Uncertain'', the instance is escalated to a fourth senior annotator.
\end{itemize}

The senior annotator has full access to:
\begin{enumerate}
    \item the original annotation bundle (source excerpt, citing context, taxonomy definitions),
    \item the three annotators' labels, and
    \item their free-text rationales.
\end{enumerate}
They then assign a final label consistent with the Dependency Precedence Rule, optionally revising the instance text if a clear typographical or formatting error is discovered. In practice, adjudication rarely required reconsidering the underlying taxonomy itself: no new error categories were requested, and all disputed cases could be resolved by clarifying which litmus question applied most directly.

Taken together, the low reliance on ``Other'' and ``Uncertain'', the substantial inter-annotator agreement, and the lack of new categories during adjudication support our central claim: the proposed five-category taxonomy is both \emph{complete} for the failure modes captured in \mb{} and \emph{operationally precise} enough to be applied reproducibly across disciplines.

\section{MisciteBench Construction Details}

\subsection{Source Selection Rationale}
\label{app:source_selection_rationale}

Our goal in constructing \mb{} is to obtain a contamination-controlled, cross-disciplinary benchmark that (i) represents the \emph{actual} scientific articles that structure their fields and (ii) admits rich downstream citation structure for the Evidence Committee in ICSV. This motivates our choice of ``most-cited article in the JIF-leading journal per JCR subject category'' rather than random sampling.

Concretely, for each of the 254 Clarivate JCR subject categories, we perform the following steps:

\begin{enumerate}
    \item \textbf{Journal selection.} We identify the 2024 Journal Impact Factor (JIF) leader within the subject category. When multiple journals share identical JIF up to three decimal places, we break ties by (i) total citable items and then (ii) alphabetical order of journal title.

    \item \textbf{Article-level selection.} Within the JIF-leading journal, we consider all ``research article''-type items published in 2024--2025.\footnote{We exclude editorials, letters, corrigenda, and news-and-views pieces by filtering on publisher-provided article-type metadata and, when ambiguous, by manual inspection.} Among these, we select the most-cited paper according to citation counts as of 2025-11-30. Citations are computed within the same JCR subject category to avoid advantaging articles that cross over into much larger neighboring fields.

    \item \textbf{Eligibility checks.} We discard candidate sources that
    (a) are not primarily original research reports (e.g., survey/overview/review-type articles, tutorials, conference overviews, and other narrative syntheses),
    (b) are direct translations or republications of older work, or
    (c) have fewer than 5 downstream citations recorded in curated citation indexes (required for ICSV to assemble a non-trivial Evidence Committee).
\end{enumerate}

This procedure offers three advantages. First, it guarantees that each source paper is a \emph{field-central} object around which substantial citation behavior has already accumulated, rather than a marginal or rarely cited work. Second, by choosing within each JCR category, we avoid overrepresenting large or fast-growing fields and maintain coverage across 21 high-level disciplines. Third, anchoring the time window in 2024--2025 maximizes exposure to contemporary citation practices and reduces the risk that frontier LLMs have memorized the full text, which is critical for the knowledge-blank protocol in Appendix~\ref{app:factcheck}.

\subsection{Knowledge-Blank Cleanroom Fact-Checking Protocol}
\label{app:factcheck}

The knowledge-blank ``cleanroom'' protocol is designed to ensure that evaluated models cannot rely on parametric memory of source papers in \mb{}, but must instead reason over the documents they are given at evaluation time. Intuitively, a paper enters \mb{} only if a broad panel of strong models collectively ``fail'' to recognize its internal details, even when explicitly prompted to recall them.

\paragraph{Fact-check probe synthesis via LRM + human curation.}

For each candidate source paper, we construct $N = 10$ \emph{forensic fact-check questions} such that answering them correctly requires access to the main body or appendices and cannot be solved from title, authors, venue, year, or general background knowledge alone. We follow these design guidelines:

\begin{itemize}
    \item Questions target \emph{specific}, non-obvious details: numerical results, ablation outcomes, limitations, cross-subgroup comparisons, key hyperparameters, or subtle qualitative caveats.
    \item Each question has a deterministic, single ground-truth answer (a number, short phrase, or Boolean) so that correctness is well-defined.
    \item The answer is supported by a single gold snippet that a human auditor can locate in the full text in a short time.
    \item Questions are balanced in difficulty, mixing relatively simple pattern-completion probes (e.g., filling in a key number) with more intricate checks that hinge on fine-grained details.
    \item All questions must be grounded in the \emph{content of the source paper itself}, not in any paper that the source cites.
\end{itemize}

To construct these probes, we first apply a Large Reasoning Model (LRM; \texttt{gemini-2.5-pro}) to the \emph{full text} of the candidate paper, including appendices. The LRM is instructed to read the paper sentence by sentence, perform deep reasoning to understand every detail, and then propose 15 candidate fact-check questions together with their canonical answers and supporting snippets. The exact prompt we use is:

\begin{codebox}[Fact-check question synthesis prompt]{}
You will be given the full text of a scientific paper (including any appendices).
Read the paper carefully, line by line, and use very deep reasoning to ensure
that you understand every detail of its content.

Your task is to design 15 fact-check (FC) questions about this paper that can be used to test whether a large language model has memorized the paper as training
data and truly knows its internal details.

Requirements:

1. Each FC question must have a single, deterministic ground-truth answer.
   The answer must not be vague, subjective, or open-ended. It should be short
   (e.g., a specific number, phrase, or Boolean) so that correctness can be
   checked reliably.

2. The FC questions must probe details that are specific to this paper.
   Avoid anything that could be guessed from common sense or generic background
   knowledge. Focus on paper-exclusive details such as exact numbers, specific
   experimental configurations, particular subsets, ablation outcomes, or
   precise verbal caveats.

3. The set of 15 FC questions should cover a balanced range of difficulty:
   - some easier checks (e.g., completing a key number in a sentence),
   - some medium-difficulty checks,
   - some harder checks that require integrating information across sections
     (e.g., main text + appendix).
   All questions must be strictly faithful to the paper; do not introduce
   hallucinated content.

4. Prefer high-memorization cues: if this paper had been used as part of a
   training set, a model that memorized it should be able to answer these
   questions accurately.

5. Very important: all FC questions must be derived from the content of this
   paper itself. Do NOT base any question on the content of other papers cited
   in the references.

Output format:

Return the 15 FC questions in standard CSV form with the following columns:
- "Paper Title": the exact title of this paper in English.
- "FC Question": the fact-check question in English.
- "Standard Answer": the unique correct answer in English.
- "Original Text": the exact snippet from the paper where the answer appears.

For any mathematical expressions in "Standard Answer" or "Original Text",
use inline LaTeX with $...$ and ensure that the formulas are completely
correct and compilable.

All content in the CSV must be in precise English.

Paper full text:
[INSERT PAPER HERE]
\end{codebox}

From the 15 LRM-generated candidates, human annotators then select and edit a final set of $N = 10$ questions per paper. During curation, annotators (i) verify that each question truly requires full-text access, (ii) remove any probe that could plausibly be answered from metadata or general domain knowledge, and (iii) correct the \texttt{Standard Answer} and \texttt{Original Text} fields when necessary.

\paragraph{Gating models and prompt.}

We probe a fixed, heterogeneous panel of frontier and strong open models:
\texttt{gpt-4o-2024-08-06},
\texttt{o4-mini-2025-04-06},
\texttt{gpt-5-2025-08-07},
\texttt{claude-sonnet-4-20250514},
\texttt{claude-opus-4-20250514},
\texttt{gpt-oss-120b},
\texttt{gpt-oss-20b},
\texttt{Nemotron 3 Nano (30B A3B)},
\texttt{Llama-3.3 Nemotron Super 49B v1},
\texttt{Qwen3-235B-A22B-Thinking-2507},
\texttt{Qwen3 VL 32B Thinking},
\texttt{Qwen3 VL 8B Thinking},
\texttt{Ministral 3 (14B Reasoning 2512)},
\texttt{Magistral Medium},
\texttt{Deepseek-V3.2 (Thinking)},
\texttt{DeepSeek-R1-0528},
\texttt{Deepseek R1 Distill Qwen 32B},
\texttt{Deepseek R1 Distill Qwen 14B},
\texttt{Llama 3.1 405B Instruct}, and
\texttt{Llama 3.3 70B Instruct}.

For each model, we deliberately restrict access to \emph{metadata only}: title, author list, journal/venue, and publication year. The abstract and full text are \emph{never} provided to the gating models. We explicitly disable tool usage and web browsing.

Each model receives, for each question, the following fact-check gate prompt.

\begin{codebox}[System Prompt for Fact-Check Gate]{}
You are evaluating whether you already know the internal contents of a scientific article from your training data.

You will see ONLY the article's title, authors, journal/venue, and year.
You will NOT see the abstract or full text.

Then you will be asked a highly specific factual question about the article.

Very important:
- Do NOT guess.
- If you are NOT certain of the exact answer based on what you already know
  about this paper from pre-training, answer with the single token: UNKNOWN.
- Only answer with a concrete value if you are sure it exactly matches what
  appears in the original paper.
\end{codebox}

\begin{codebox}[User Prompt for Fact-Check Gate]{}
Article metadata
Title:   [TITLE]
Authors: [AUTHORS]
Journal: [VENUE]
Year:    [YEAR]

Fact-check question
[QUESTION_i]

Instructions
If you are certain you know the exact answer from your internal knowledge
of this paper, answer with a short phrase or number.

Otherwise, answer with the single token: UNKNOWN.
\end{codebox}

\paragraph{Automatic grading and acceptance criterion.}

For each question, we store the canonical gold answer $a_i$. Let $r_i^{(m)}$ denote model $m$'s response to question $i$.

\begin{itemize}
    \item \textbf{Numeric answers.} If $a_i$ is numeric, we parse $r_i^{(m)}$ as a real number whenever possible and count it as correct if
    \[
        \frac{|r_i^{(m)} - a_i|}{\max(1, |a_i|)} < 0.01.
    \]

    \item \textbf{Textual answers.} If $a_i$ is textual, we do not rely on brittle string equality. Instead, we use an independent LLM grader (\texttt{gpt-4o-2024-08-06}) that receives both the gold answer and the model's response and decides whether they are \emph{semantically equivalent}. The grading prompt is:

\begin{codebox}[Grading Prompt]{}
You are grading whether two short answers express the same factual content.

Gold answer:    "[GOLD_ANSWER]"
Model answer:   "[MODEL_ANSWER]"

If, ignoring superficial differences in wording, the two answers refer to the
same specific fact with the same level of specificity, respond with YES.
Otherwise respond with NO.

Respond with exactly one token: YES or NO.
\end{codebox}

    A textual response $r_i^{(m)}$ is marked correct if and only if the grader returns \texttt{YES}.
\end{itemize}

A candidate paper is \emph{admitted} into \mb{} if and only if every model in the panel fails all probes, i.e.,
\[
    \forall m \in \mathcal{M}, \quad \sum_{i=1}^{N} \mathbf{1}\big[r_i^{(m)} \text{ is correct}\big] = 0.
\]
If any model answers at least one probe correctly, we discard the paper and move to the next most-cited article in the same journal and time window, repeating the procedure until a clean paper is found or the candidate pool is exhausted.

\subsection{Dual-Tier Adversarial Miscitation Generation and Validation}
\label{app:verification}

This section details how we generate and validate adversarial miscitations grounded in the five-category taxonomy described in Section~\ref{subsec:taxonomy}. The goal is to create instances that are (i) realistic, (ii) taxonomy-pure (each miscitation belongs unambiguously to one category), and (iii) calibrated in difficulty, ranging from surface-level errors to expert-level traps.

\paragraph{LRM generation prompt and output schema.}

Given a source paper $B$, we provide a Large Reasoning Model (\texttt{gemini-2.5-pro}) with the \emph{full} paper, including all main sections and appendices, together with the detailed definitions of the five miscitation categories. The model is instructed to first perform extremely deep reading and then, for each category, synthesize both moderate and very difficult miscitations.

The core generation prompt (simplified for presentation) is:

\begin{codebox}[Prompt to LRM for miscitation synthesis.]{}
We are constructing a dataset of diverse miscitation examples to evaluate
automatic miscitation detection systems.

You will be given the FULL TEXT of a source paper (including appendices) and
the definitions of 5 miscitation categories. First, read the paper slowly and
carefully, sentence by sentence, and use very deep reasoning to ensure that
you understand its methods, results, limitations, and nuanced details.

Then, for EACH of the 5 miscitation categories, you must create 5 miscitation
examples that (incorrectly) cite this paper:

- 3 examples should be single sentences or short paragraphs that are clearly
  wrong for a careful reader who deeply understands the paper, but not so
  trivial that they can be spotted at a glance.

- 2 examples should be long, complex sentences or slightly longer paragraphs
  that are VERY difficult to detect as wrong. These "deep" miscitations should
  rely on subtle, easily overlooked details of the paper, such that even domain
  experts or the original authors would need to think carefully before
  identifying the error.

Additional constraints:

- Different scenarios under the same category should be diverse. Cover as many
  distinct ways of committing that type of miscitation as possible.

- Different categories must remain well separated. Each designed example should
  clearly belong to exactly ONE miscitation category under the taxonomy, with
  no ambiguity or overlap.

- For EVERY miscitation example, you must also provide:
  * Explanation: a clear English explanation of why this is a miscitation and
    how it instantiates the target category.
  * Correct Statement: a corrected citing sentence or short passage that would
    describe the source paper accurately.
  * Original Text: the exact snippet(s) from the source paper that grounds
    your judgment (i.e., what the paper actually says).

- Read the source paper deeply enough to avoid being misled by possible
  typesetting or formatting errors.

- Very important: miscitation content must be about THIS source paper. Do NOT
  base any miscitation on the content of other papers cited in its references.

Output format:

Return all 25 designed miscitations (5 categories * 5 examples) in CSV form
with the following columns:

- "Miscitation":        the incorrect citing sentence or short paragraph.
- "Explanation":        why this is a miscitation, in clear English.
- "Correct Statement":  a corrected version that would cite the paper properly.
- "Original Text":      the supporting snippet(s) from the source paper.
- "Miscite Type":       one of {Citation Validity Error,
                         Content Misrepresentation Error,
                         Scope Extrapolation Error,
                         Evidence Characterization Error,
                         Attribution & Traceability Error}.
- "Difficulties":       one of {SURFACE, DEEP}, where SURFACE indicates a
                         moderate, more local error, and DEEP indicates a
                         very difficult, globally subtle error.

All content must be in precise English. Ensure that each example strictly
matches its assigned miscitation type and difficulty level.

Source paper full text:
[INSERT PAPER HERE]

Miscitation taxonomy:
[INSERT 5 CATEGORY DEFINITIONS HERE]
\end{codebox}

This prompt yields, for each source paper, 25 candidate miscitation rows covering the full taxonomy:
three \texttt{SURFACE} and two \texttt{DEEP} instances per category, with explicit fields
\texttt{Miscitation}, \texttt{Explanation}, \texttt{Correct Statement}, \texttt{Original Text},
\texttt{Miscite Type}, and \texttt{Difficulties}. We perform schema validation and reject any row where
the generated \texttt{Miscite Type} is inconsistent with the textual description in \texttt{Explanation}.

\paragraph{Post-processing and CSV representation.}

The raw LRM outputs are parsed into a standardized CSV representation. We normalize category labels to the five taxonomy names, collapse minor variations in difficulty tags into \texttt{SURFACE} vs.\ \texttt{DEEP}, and enforce basic well-formedness constraints (non-empty fields, length limits, and valid category labels). Instances that appear to mix multiple error types (e.g., both misattribution and content distortion) are flagged for later removal or manual repair.

\paragraph{Independent LRM consistency check.}

Each candidate miscitation is then re-evaluated by an \emph{independent} LRM with a different architecture and training history (\texttt{gpt-5.1-Extended-Thinking}). For each row, the independent model receives:

\begin{itemize}
    \item the full source paper,
    \item the candidate \texttt{Miscitation} sentence/paragraph,
    \item the proposed \texttt{Miscite Type} and \texttt{Difficulties},
    \item the \texttt{Original Text} snippet, and
    \item the original \texttt{Explanation} and \texttt{Correct Statement}.
\end{itemize}

The independent LRM is asked to (i) decide whether the candidate is in fact a miscitation of the source, (ii) assign the most appropriate taxonomy category, and (iii) provide its own free-form explanation of the error and a corrected citing statement.

We then apply two automatic filters:

\begin{itemize}
    \item \textbf{Category agreement.} If the independent model assigns a different primary miscitation type than the original \texttt{Miscite Type}, the instance is flagged for removal or manual repair.
    \item \textbf{Explanation alignment.} We again use \texttt{gpt-4o-2024-08-06} as a semantic grader to compare the independent model's explanation to the original \texttt{Explanation}. Only instances where the two explanations are judged to be semantically equivalent (same mechanism, same primary error) are retained.
\end{itemize}

Examples that fail either check are either edited and re-validated or dropped entirely.

\paragraph{Human expert validation.}

Finally, we subject every source paper and its 25 candidate miscitations to rigorous human review by domain experts. We recruit a panel of PhD students, researchers, and senior practitioners spanning all 21 high-level disciplines covered by the 254 JCR subject categories. Collectively, their expertise covers the full set of subfields represented in \mb{}.

For each source paper, we assign 3--4 experts whose research area matches the corresponding high-level discipline (e.g., Clinical Medicine, Computer Science, Social Sciences). Each expert is given the full source paper and the 25 miscitation rows and asked to perform cross-validation under the following criteria:

\begin{enumerate}
    \item \textbf{Is it truly a miscitation?}  
    Does the \texttt{Miscitation} sentence or paragraph in fact misrepresent the source paper, and does the stated \texttt{Explanation} accurately and fully capture the reason it is wrong?

    \item \textbf{Taxonomy alignment.}  
    Given the definitions in our five-category taxonomy, does the assigned \texttt{Miscite Type} correspond to the \emph{primary} error mechanism? Could a reasonable annotator confidently place this instance in exactly this category?

    \item \textbf{Evidence grounding.}  
    Does the \texttt{Original Text} field correctly and sufficiently point to the span(s) in the source paper that justify the error diagnosis (and, for corrected statements, their validity)?

    \item \textbf{Corrected statement accuracy.}  
    Is the \texttt{Correct Statement} factually consistent with the source paper and free from new miscitation issues?
\end{enumerate}

A miscitation instance is retained in \mb{} only if all assigned experts agree that it satisfies the criteria above. If any expert disagrees on miscitation status, taxonomy label, evidence grounding, or corrected-statement accuracy, the instance is either revised and re-submitted to the same validation protocol or discarded.

After this multi-stage pipeline---LRM generation with deep reading, independent cross-model consistency checking, and stringent domain-expert cross-validation---we obtain the final set of 254 source papers $\times$ 5 categories $\times$ 5 instances per category, for a total of 6{,}350 high-quality miscitation cases used in \mb{}.

\section{Document Parsing and Citation Mapping Details}
\label{app:dpcm}

The Document Parser and Citation Mapper (DPCM) normalizes heterogeneous inputs
(\LaTeX{}, XML/HTML, and PDFs) into a single hierarchical Markdown
representation. This representation preserves the discourse skeleton of the
paper—headings, paragraphs, math, figures/tables, and citation anchors—so that
downstream verification always operates on a traceable, format-agnostic
substrate. This appendix details (i) markup-based normalization,
(ii) PDF parsing via image-based layout serialization,
(iii) the Extraction Verifier, and (iv) citation mapping.

\subsection{Markup-Based Normalization}

For \LaTeX{} and XML/HTML inputs, DPCM constructs an intermediate abstract
syntax tree (AST) and then projects it into Markdown while deliberately
retaining only semantics that matter for citation auditing.

\paragraph{Macro handling and structural projection.}
We expand standard inline macros such as \verb|\emph|, \verb|\textbf|, and
\verb|\textit| where they affect emphasis, and prune purely presentational
directives (spacing, color, line breaks) that do not change logical structure.
Sectioning commands (\verb|\section|, \verb|\subsection|, etc.) are converted
into Markdown headings (\verb|#|, \verb|##|, \verb|###|), while lists, display
math, and inline math are serialized as Markdown lists and \verb|$...$| / \verb|$$...$$|
blocks. Cross-references (\verb|\ref|, \verb|\eqref|, etc.) are preserved as
literal text so that equation, figure, and table numbering can later be checked
for consistency.

\paragraph{Citation rendering and anchoring.}
Citation commands in markup (\verb|\cite|, \verb|\citep|, \verb|\citet| and
style-specific variants) are processed in two coupled layers:

\begin{enumerate}
  \item \emph{Surface rendering.}
  Using the document’s bibliographic style configuration (e.g., numeric,
  author–year/APA, Chicago), we re-render each citation into the same surface
  form that a compiled PDF would produce: for example,
  \verb|\citep{Smith2020,Brown2019}| becomes ``(Smith, 2020; Brown \& Brown, 2019)''
  under APA-style settings or ``[12, 19]'' under a numeric style.
  As a result, the Markdown closely mirrors the citing text as seen by human
  readers, which is crucial for later LLM-based reasoning.
  \item \emph{Stable internal anchors.}
  In parallel, DPCM attaches an internal anchor to each rendered citation
  that records the originating citation keys and their order. These anchors
  are \emph{not} shown in the surface Markdown, but they form a lossless
  mapping from each inline citation occurrence back to the corresponding
  bibliography entries. Citation mapping (Section~\ref{app:dpcm_citation_mapping})
  later consumes these anchors to build sentence-level citation graphs,
  while ACSV and ICSV operate purely on the human-readable surface text.
\end{enumerate}

The result of markup normalization is a hierarchical Markdown document that
faithfully reflects the compiled reading experience, while still exposing a
machine-tractable citation backbone.

\subsection{PDF Parsing via Image-Based Layout Serialization}
\label{app:dpcm_pdf}

Text-based PDF extraction often yields broken reading order, missing math, and
unreliable treatment of figures and tables. For PDFs, DPCM therefore uses an
explicit image-based pipeline that delegates layout reconstruction to a
multimodal model and then repairs cross-page boundaries.

\subsubsection{Page rendering and page-level transcription}

Each PDF is first converted into one raster image per page at a fixed
resolution of 300 DPI using \texttt{pdf2image}. The parser does not rely on
embedded PDF text; instead, each page image is passed directly to
\texttt{gpt-4o-2024-08-06} together with a concise system prompt that specifies
the Markdown serialization rules.

The system prompt used for page-level transcription is:

\begin{codebox}[]{}
Please accurately convert the main body text in the image into markdown.
Use $...$ for inline formulas and $$...$$ for display formulas.

Requirements:
1. Preserve the original wording exactly. Do NOT paraphrase, add, or omit text.
2. Extract only main body text.
   - Ignore illustrations and graphical elements.
   - Keep captions whose text starts with "Figure", "Table", "Fig.", or "Tab.".
3. If the image conta`ins tables, convert them faithfully into markdown tables.
4. Remove headers, footers, page numbers, and sidebars that are not part of
   the main body.
5. Do not call any external tools; rely only on your visual understanding.
6. If the image contains no readable text, output: null
\end{codebox}

The model returns Markdown for the page; if it wraps the content inside a code
fence, we strip the fence and retain only the inner Markdown. This gives us a
sequence of per-page Markdown fragments that already approximate the correct
reading order, with headings, paragraphs, displayed math, inline math, figure
and table captions, and inline citations rendered as they appear on the page.

\subsubsection{Cross-page boundary repair}

Page-level OCR and transcription inevitably break some sentences across page
boundaries. DPCM therefore applies a dedicated cross-page repair pass before
any structural verification or citation mapping.

We maintain two parallel views of each page:
(i) the original page-level Markdown output, and
(ii) a tagged version used for debugging, in which we mark potentially
incomplete beginnings or endings using lightweight heuristics.

\paragraph{Incomplete head/tail tagging.}
For each page, we inspect the first and last non-empty lines. A line is
considered a \emph{complete} tail if it ends with sentence-final punctuation
(e.g., ``.'', ``!'', ``?'' or their CJK counterparts), closes a math block
(\verb|$| or \verb|$$|), ends with a citation or reference pattern (e.g.,
``[12]''), or is a plausible standalone heading or caption. We consider the
first content line to be a \emph{complete} start if it looks like the beginning
of a new sentence or block (e.g., capitalized heading, list item, or math
environment) rather than a continuation.

If the tail of a page fails these checks, we tag it with
\verb|<INCOMPLETE_END_Pn>|; analogously, if the head of the next page looks
like a continuation (lowercase start, conjunctions such as ``and'', ``but'',
``however'', leading bracket, etc.), we tag it with \verb|<INCOMPLETE_START_Pn>|.
These tags do not change the content but mark candidate boundaries for repair.

\paragraph{Heuristic merging and hyphen repair.}
We then merge per-page texts into a single document using a boundary-aware
procedure. For each adjacent page pair, we:

\begin{itemize}
  \item Split each page into paragraphs separated by blank lines.
  \item Separate figure/table captions from main text, so that caption-only
  paragraphs are appended without being merged into running prose.
  \item Examine the last main-text paragraph of page $n$ and the first
  main-text paragraph of page $n{+}1$.
\end{itemize}

If the tail paragraph is incomplete and the head paragraph does not resemble a
new heading or section, we attempt to repair the boundary:

\begin{enumerate}
  \item \emph{Hyphenated words.} If the tail ends with a hyphenated fragment
  (e.g., ``multi-'' at the end of a line) and the head begins with alphabetic
  characters, we merge them into a single word (e.g., ``multi-agent''), keeping
  a single hyphen when appropriate.
  \item \emph{Direct continuation.} If the head starts with a lowercase letter,
  discourse connective (\emph{and, but, however, therefore, which, that,
  because}, etc.), or an opening bracket, we join the two paragraphs with a
  space, treating them as a single sentence that crossed the page break.
\end{enumerate}

If neither heuristic confidently applies but the tail paragraph is clearly
incomplete and not a caption, we fall back to a minimal LLM-based boundary
repair step that edits only the ambiguous junction.

\paragraph{LLM-based boundary repair.}
For hard cases, we send the last paragraph of page $n$ and the first paragraph
of page $n{+}1$ to \texttt{gpt-4o-2024-08-06} with a dedicated system prompt:

\begin{codebox}[]{}
You repair cross-page sentence breaks in markdown text extracted from a
scientific PDF. You are given:

- PREV: the tail paragraph of page N
- NEXT: the head paragraph of page N+1

They may contain:
- a sentence split across the page break,
- duplicated fragments, or
- a word split by a hyphen.

Your task:
- Minimally fix the boundary so that the text reads as a single continuous
  document.
- Do NOT invent new content or change meaning.
- Do NOT drop valid content except for exact duplicates.
- You may:
  - move tokens between PREV and NEXT,
  - remove duplicated fragments,
  - join hyphenated words split across the boundary.

Output exactly in this format:

PREV_FIXED:
<fixed text for the previous-page tail>
---
NEXT_FIXED:
<fixed text for the next-page head>
\end{codebox}

We parse \verb|PREV_FIXED| and \verb|NEXT_FIXED| from the model output and
adopt them only if the boundary has changed in a way that improves continuity
(e.g., duplication removed, sentence completed). Otherwise, we keep the
original paragraphs. This design ensures that the LLM operates as a local
repair operator, not as a free-form rewriter of the document.

After iterating over all boundaries, we remove debug tags
\verb|<INCOMPLETE_START_Pn>| and \verb|<INCOMPLETE_END_Pn>| and normalize
whitespace, yielding a single merged Markdown file that reflects the visual
layout and reading order of the PDF while repairing cross-page sentence breaks.

\subsection{Extraction Verifier}
\label{app:dpcm_verifier}

Once the merged Markdown is available, the Extraction Verifier performs a
multi-level audit to detect structural and bibliographic inconsistencies before
citation reasoning. The verifier combines deterministic checks with targeted
LLM-based inspection.

\paragraph{Level 1: deterministic structural checks.}
We parse the Markdown into a lightweight document tree and enforce basic
invariants:

\begin{itemize}
  \item \emph{Heading progression.} Heading levels must not jump by more than
  one level (e.g., \verb|#| $\rightarrow$ \verb|##| $\rightarrow$ \verb|###| is
  allowed; \verb|#| $\rightarrow$ \verb|###| is flagged). Abrupt regressions
  (e.g., \verb|####| after \verb|#|) are similarly flagged.
  \item \emph{Equation and float numbering.} Equation labels (``(1)'',
  ``(2)'', \dots) and figure/table numbers (``Figure 3'', ``Table 2'') are
  checked for monotone, mostly contiguous sequences. Small gaps are tolerated,
  but long stretches of missing or duplicated numbers trigger a warning.
  \item \emph{Citation index sequences.} For numeric styles, we extract all
  integers appearing in citation-like patterns (e.g., ``[3]'', ``[3, 5, 9]'',
  ``[3--5]'') and verify that the global sequence is roughly monotone and
  dense. Large gaps (e.g., no references between [5] and [20]) or frequent
  out-of-order jumps signal potential parsing failures, such as skipped
  bibliography segments or mis-ordered text blocks.
\end{itemize}

For each violation, the verifier records the surrounding lines and the page
range where the anomaly occurs.

\paragraph{Level 2: localized re-parsing.}
When Level~1 identifies a likely structural discontinuity that aligns with a
small set of pages (e.g., a break in equation numbers between pages 7 and 8),
we re-run the PDF-to-Markdown conversion only on those pages. The re-parse uses
the same 300~DPI images and system prompt as in
Section~\ref{app:dpcm_pdf}, but the new output is compared against the original
to detect missing lines, duplicated blocks, or altered reading order. If the
re-parsed fragment resolves the anomaly (e.g., restores a missing equation or
bibliography block), we splice it into the global document; otherwise, we keep
the original and proceed to Level~3.

\paragraph{Level 3: LLM-based semantic audit.}
For residual suspicious regions where deterministic rules cannot confidently
decide (e.g., text density drops abruptly without clear structural markers), we
invoke a light semantic audit. We provide \texttt{gpt-4o-2024-08-06} with the
Markdown snippet around the anomaly and a succinct description of the expected
structure, asking whether the snippet appears truncated, duplicated, or
out-of-order. A representative prompt is:

\begin{codebox}[]{}
You are auditing markdown extracted from a scientific paper.

You are given a short markdown segment and a brief description of the
surrounding document structure (e.g., "between Section 3.2 and Section 3.3").

Decide whether the segment shows signs of extraction error:
- missing lines or sentences,
- duplicated paragraphs,
- obvious reading-order errors (e.g., a caption in the middle of a sentence).

Answer with one of:
- OK
- SUSPICIOUS_MISSING
- SUSPICIOUS_DUPLICATE
- SUSPICIOUS_ORDER

Then, in one short sentence, explain your choice using only evidence that is
visible in the provided markdown.
\end{codebox}

Segments labeled as suspicious are either re-parsed again with tighter crops or
flagged as low-confidence regions for downstream modules (ACSV/ICSV) and, if
needed, for manual inspection. In practice, this three-level verifier reduces
gross extraction failures (e.g., missing sections, mis-ordered columns) while
keeping the pipeline predominantly deterministic and auditable.

\subsection{Citation Mapping}
\label{app:dpcm_citation_mapping}

Given a structurally verified Markdown document, the citation mapping component
constructs a sentence-level citation graph that is consumed by CSAC, ACSV, and
ICSV. The goal is to map each inline citation span in the body text to one or
more normalized bibliography entries, while preserving the original surface
style.

\paragraph{Inline citation detection.}
We first identify citation spans in the Markdown using style-agnostic patterns.
The detector considers three families:

\begin{itemize}
  \item \emph{Numeric citations}, such as ``[12]'', ``[3, 5, 9]'',
  ``[3--5]'', and inline variants like ``(see also [7])''.
  \item \emph{Author–year citations}, such as ``(Smith, 2020)'',
  ``Smith (2020)'', and grouped forms like
  ``(Smith, 2020; Brown \& Lee, 2019)''.
  \item \emph{Footnote-based references}, where the citation marker appears as
  a superscript or bracketed index in the main text and the full reference
  appears in a footnote block.
\end{itemize}

By scanning the entire document, we infer the dominant citation style (numeric
vs.\ author–year vs.\ note-based) from frequency statistics and pattern
coverage, and adapt parsing rules accordingly. Each detected span is associated
with its containing sentence (using punctuation-based segmentation with
list/heading safeguards), yielding a preliminary mapping from sentences to
unnormalized citation strings.

\paragraph{Bibliography parsing.}
We then isolate the bibliography section (or sections) by detecting headings
such as ``References'', ``Bibliography'', or style-specific variants. Each
bibliographic entry is parsed into a normalized record containing at minimum:
canonical author list (last names and initials), publication year,
title tokens, venue/journal name, and any explicit identifiers (DOI, arXiv ID,
PubMed ID). For markup-based inputs, we prefer the original \verb|\bibitem|
or Bib\TeX{} metadata; for PDF-only inputs, we rely on typography cues (hanging
indentation, numbering, bullet markers) combined with pattern-based parsing.

\paragraph{Citation-to-entry alignment.}
Finally, we align inline citation spans to bibliography entries:

\begin{itemize}
  \item \emph{Markup inputs.} When the document originates from \LaTeX{} or
  XML/HTML, we use the internal anchors described earlier to directly map each
  inline citation occurrence to a set of bibliography records, with no string
  matching required. This mapping is exact and order-preserving.
  \item \emph{PDF-only inputs, numeric style.} We map numeric indices in
  citations (e.g., ``[7]'' or ``[3--5]'') to positions in the parsed
  bibliography list, taking into account style-specific conventions (e.g.,
  whether numbering restarts in supplements). Ranges and grouped citations are
  expanded into individual edges (e.g., ``[3--5]'' yields links to 3, 4, and 5).
  \item \emph{PDF-only inputs, author–year style.} For each citation string, we
  extract last names and year(s), then compute a similarity score between this
  tuple and each candidate bibliography entry, based on overlap of normalized
  last names, year equality, and title token similarity. The highest-scoring
  candidate above a threshold is selected; ties and sub-threshold cases are
  flagged as ambiguous.
\end{itemize}

In ambiguous or noisy cases (e.g., partial author lists, missing years), we use
a lightweight LLM-assisted disambiguation step: we provide the surface citation
string and a small set of candidate bibliography entries and ask the model to
select the best match or to abstain if none is appropriate. This step is
constrained to choose from explicit candidates and does not fabricate new
references.

The final output of citation mapping is, for each sentence including intext citation(s) in the document, a set of resolved citation edges pointing to normalized metadata records (title, authors, year, venue, DOI). This structure is the entry point for CSAC, which determines source accessibility, and for ACSV/ICSV, which perform taxonomy-aligned miscitation verification on top of a fully traceable citation graph.

\subsection{Source Accessibility Classification (CSAC)}

CSAC operates on the normalized metadata produced by the Citation Mapper and decides how each cited source will be verified. It first attempts to resolve a DOI; when a DOI is present, CSAC queries publisher and venue APIs for full text, falling back to metadata-only records when full text is restricted. When no DOI is available, CSAC constructs a metadata query from the title, author list, and year and searches publisher indices and curated open-access repositories for candidate matches. In all cases, any retrieved open-access surrogate must pass strict equivalence checks based on title similarity, author overlap, abstract content, and, when available, reference-list signatures; only then is it treated as the same scientific work as the cited source.

Based on these retrieval outcomes, CSAC assigns each reference to one of three regimes. References with verified full text—either the publisher version or an accepted open-access surrogate—are marked as accessible and routed to ACSV. References with stable metadata but no full-text access are marked as inaccessible and routed to ICSV together with their metadata snapshot (title, authors, abstract, venue, year, and any available identifiers). References for which neither DOI resolution nor repository search yields a plausible match are labeled Ghost Citations, providing a direct signal for Attribution \& Traceability errors and preventing \ba\ from fabricating surrogate documents or hallucinated rationales for non-existent sources.

\section{ICSV Implementation Details}
\label{app:icsv}

This appendix documents the concrete implementation of the \textbf{Inaccessible Cited Source Verifier (ICSV)} and its \textbf{Evidence Committee} mechanism. ICSV targets the strict paywall regime: the full text of the cited source $B$ is unavailable, so verification must be grounded in \emph{auditable downstream evidence} rather than speculative reconstruction. We therefore (i) extract a \emph{claim-preserving paraphrase} of what the citing paper $A$ attributes to $B$, (ii) extract analogous attributions to $B$ from multiple open-access downstream citers, (iii) organize these attributions into coherent aspects via LLM-based semantic clustering, (iv) assign field-normalized influence weights across heterogeneous venue types (journal / conference / preprint), and (v) compute a reliability-aware consensus verdict with calibrated confidence and principled abstention.

Unless otherwise noted, all ICSV LLM calls use \texttt{gpt-4o-2024-08-06}. We use \texttt{gpt-4o-2025-08-06} only for the semantic clustering step (Section~\ref{app:icsv:cluster}) to improve partition stability on long claim lists.

\vspace{0.5em}
\subsection{Downstream Committee Retrieval and Witness Verification}
\label{app:icsv:retrieval}

Given a paywalled cited source $B$, CSAC provides a resolved metadata snapshot (DOI when available; otherwise title, author list, venue, year). ICSV then constructs an \emph{open-access witness set} $C_{\text{open}}$ of downstream citers that reference $B$ and have retrievable full text.

\paragraph{Primary citation-graph retrieval.}
We query a curated open citation index (OpenAlex as primary; Crossref as DOI/metadata fallback) to enumerate works that cite $B$. We de-duplicate by DOI and canonical title normalization (lowercasing, punctuation stripping, whitespace collapse) and discard non-scholarly records (e.g., editorial notes) using venue/type metadata when available.

\paragraph{Open-access eligibility and full-text acquisition.}
A candidate citer $p$ enters $C_{\text{open}}$ only if (i) a full-text URL is available (publisher OA, institutional repository, or vetted preprint server), and (ii) the downloaded full text can be parsed by DPCM into structured Markdown with intact citation anchors.

\paragraph{Witness validity check (must explicitly cite $B$ in-text).}
For each candidate citer $p$, we verify that $p$ contains at least one \emph{explicit in-text citation mention} to $B$. We accept a mention if any of the following match robustly:
(i) DOI match, (ii) high-similarity title string match, or (iii) bibliography-entry match followed by an in-text anchor pointing to that entry.
Candidates that do not pass this in-text witness check are discarded to prevent ``false witnesses'' created by noisy citation graphs.

\paragraph{No premature skipping.}
ICSV never ``skips'' a paywalled citation after retrieval. If witnesses are weak or contradictory, the system returns \textsc{Undecidable} with an explicit reliability diagnosis (Section~\ref{app:icsv:abstain}).

\vspace{0.5em}
\subsection{Context-Aware Citing Claim Paraphrase (from Paper $A$)}
\label{app:icsv:claimA}

The first step is to construct $c_A$: a \emph{claim-preserving paraphrase} of what $A$ attributes to $B$ at the citation site. Importantly, our implementation does \textbf{not} ask the model to ``rewrite the single main claim'' (which invites summarization or abstraction). Instead, it performs a \textbf{tight paraphrase} of the cited sentence, with one permitted transformation: \emph{resolve pronouns and implicit references into explicit mentions} while preserving all qualifiers, modality, and scope.

\paragraph{Window expansion (iterative; no skipping).}
Let $s_A$ be the in-text sentence in $A$ that cites $B$. We start from a minimal local window and expand only when necessary:
\[
W_A(r) = \text{sent}(A, i-r) \oplus \cdots \oplus \text{sent}(A, i) \oplus \cdots \oplus \text{sent}(A, i+r),
\]
where $i$ is the index of $s_A$ and $r$ is the radius (default start $r{=}1$). If the model returns \texttt{INSUFFICIENT\_CONTEXT}, we increment $r \leftarrow r+1$ and retry, continuing until a stable paraphrase is produced. We cap expansion by paragraph boundaries; if the paragraph is still insufficient, we extend to adjacent paragraphs in the same section.

\paragraph{Stability criterion.}
To avoid oscillating paraphrases under larger windows, we require \emph{two consecutive radii} to produce paraphrases that are identical after normalization (whitespace collapse; punctuation normalization) or have semantic similarity above a high threshold (measured by a sentence embedding cosine similarity). The earlier paraphrase is then fixed as $c_A$.

\textbf{Prompt template (claim-preserving paraphrase).}

\begin{codebox}[System Prompt]{}
You are an expert scientific copy-editor and verifier.
Your job is to produce a claim-preserving paraphrase of a specific sentence that cites a prior paper B. Do NOT summarize, generalize, or add new claims.
Preserve all qualifiers (e.g., "may", "suggest", "in our setting"), all scope constraints (population/task/conditions), and all numerical content.
The only required transformation is to resolve pronouns and implicit mentions into explicit referents, using the provided context.

\end{codebox}

\begin{codebox}[User Prompt]{}
[Context window W_A]
{W_A}

[Target sentence s_A that contains the in-text citation to paper B]
{s_A}

Instructions:
1) Paraphrase s_A as closely as possible (claim-preserving).
2) Resolve pronouns/implicit references using W_A (e.g., "this method" -> the named method).
3) Keep modality, qualifiers, and scope exactly faithful to s_A.
4) Do NOT mention citation markers, citation numbers, authors, or years.
5) Output ONE sentence only.
6) If the referent needed to resolve pronouns is not recoverable from W_A, output:
INSUFFICIENT_CONTEXT
\end{codebox}

\vspace{0.5em}
\subsection{Witness Claim Extraction (from Each Downstream Citer)}
\label{app:icsv:witness-claims}

ICSV extracts what each open-access witness paper $p \in C_{\text{open}}$ claims about $B$ at its explicit in-text citation sites.

\paragraph{Selecting explicit mention sites.}
For each witness $p$, we locate every sentence $s_p$ whose in-text citation anchor resolves to $B$ (via DOI/title/bibliography match as in Section~\ref{app:icsv:retrieval}). For each such $s_p$, we run the \emph{same} claim-preserving paraphrase procedure as Section~\ref{app:icsv:claimA}, producing a witness claim $c_p^{(t)}$ for mention $t$. Within a witness paper, we de-duplicate claims using high-similarity filtering to avoid overweighting repeated boilerplate mentions.

\paragraph{Outcome.}
This yields a multi-set of witness claims
\[
\mathcal{C}_B = \{(c_1,\mathrm{src}(c_1)), \ldots, (c_m,\mathrm{src}(c_m))\},
\]
where each $c_\ell$ is a claim-preserving paraphrase attributed to $B$ and $\mathrm{src}(c_\ell)$ denotes the source witness paper that produced it.

\vspace{0.5em}
\subsection{LLM-Based Semantic Clustering and Evidence Statement Distillation}
\label{app:icsv:cluster}

The goal is to organize $\mathcal{C}_B$ into coherent ``aspects'' of $B$ (e.g., method, dataset, empirical finding), then distill each aspect into a canonical evidence statement $e_j$. In our implementation, we \textbf{do not} use embedding-based agglomerative clustering, because choosing thresholds and cluster counts is brittle across fields and can merge distinct aspects of $B$ into the same cluster. Instead, we perform \textbf{direct semantic clustering with an LLM}, which is both more controllable (via explicit constraints) and more robust to domain shift.

\subsubsection{Semantic clustering prompt (LLM classifier)}
\label{app:icsv:cluster:prompt}

We use \texttt{gpt-4o-2025-08-06} to cluster witness claims into non-overlapping groups $\{G_j\}_{j=1}^k$. The prompt is designed to (i) prevent ``semantic over-merging,'' (ii) keep clusters interpretable, and (iii) enforce strict JSON output for deterministic parsing.

\begin{codebox}[System prompt (semantic clustering).]{}
You are an expert scientific librarian.
You will receive a list of short, claim-preserving sentences that different
papers attribute to a paywalled paper B.

Task: cluster these sentences into semantically coherent aspects of B.
Each cluster must represent ONE aspect (e.g., one method contribution,
one dataset, one key empirical finding, one theoretical claim).
Do NOT merge two distinct aspects just because they are topically related.

Constraints:
- Every claim must belong to exactly one cluster.
- Clusters should be as few as possible, but no cluster may contain claims
  that are substantively about different contributions/aspects.
- If a claim is vague, place it into the closest cluster only if consistent;
  otherwise create a small "misc/unclear aspect" cluster.
Return JSON only (no prose).
\end{codebox}

\begin{codebox}[User prompt.]{}
Paper B (metadata):
Title: {title_B}
Year: {year_B}
Venue: {venue_B}

Witness claims about B (each has an ID):
1) {c_1}
2) {c_2}
...
m) {c_m}

Output JSON with the following schema:
{
  "clusters": [
    {
      "cluster_id": "C1",
      "cluster_name": "short aspect name",
      "aspect_summary": "one-sentence description of the shared aspect",
      "claim_ids": [1, 7, 12]
    },
    ...
  ]
}

Rules:
- cluster_name must be <= 8 words.
- aspect_summary must be exactly one sentence.
- Do not invent content not present in the claims.
\end{codebox}

\subsubsection{Evidence statement distillation (per cluster)}
\label{app:icsv:distill}

For each cluster $G_j$, we distill a canonical evidence statement $e_j$ that represents the shared content \emph{as attributed by the community}. This step uses \texttt{gpt-4o-2024-08-06}.

\begin{codebox}[System prompt (evidence distillation).]{}

You are an expert scientific verifier.
You will see multiple claims that different papers attribute to a paywalled
paper B about the SAME aspect. Your job is to write ONE canonical evidence
statement that captures only their overlap.

Requirements:
- One sentence only.
- Include critical qualifiers (scope, conditions, uncertainty).
- Preserve numerical quantities if present and consistent.
- If claims conflict, produce a conservative statement that reflects only
  what is common, and explicitly hedge (e.g., "is reported to", "suggests").
Do NOT add new facts.

\end{codebox}

\begin{codebox}[User Prompt.]{}
Paper B (metadata):
Title: {title_B}

Cluster {cluster_id}: {cluster_name}
Claims in this cluster:
- {c_a}
- {c_b}
...

Write ONE sentence as the canonical evidence statement e_j.
\end{codebox}

\paragraph{Provenance.}
For traceability, we store for each $e_j$ the full set of contributing claim IDs and their source papers. All downstream weighting and voting operates on \emph{unique witness papers} per cluster (Section~\ref{app:icsv:influence}), preventing repeated mentions from the same paper from inflating support.

\vspace{0.5em}
\subsection{Field-Normalized Influence Weights Across Journals, Conferences, and Preprints}
\label{app:icsv:influence}

Each evidence statement is only as credible as its witnesses. However, raw citations and venue prestige vary drastically across fields and publication years, and venue types differ (journals vs.\ conferences vs.\ preprints). We therefore compute a unified influence score $\mathcal{I}(p)$ for each witness paper $p$, combining (i) \textbf{paper-level influence} within its field-year and (ii) \textbf{venue-level standing} within its field.

\subsubsection{Paper-level citation percentile (all venue types)}
Let $\mathrm{Cite}(p)$ be the paper-level citation count from a citation index snapshot. We compute a field-year normalized percentile:
\[
C_{\text{norm}}(p) = \mathrm{Rank}_{\%}\big(\mathrm{Cite}(p)\mid \mathrm{Field}(p), \mathrm{Year}(p)\big).
\]
To reduce heavy-tail instability, we winsorize citation counts within each field-year at the 99th percentile before ranking.

\subsubsection{Venue-level standing percentile (type-aware but unified output)}
We define a venue standing percentile $V_{\text{norm}}(p)\in[0,1]$ based on the venue type:

\paragraph{Journals.}
If $p$ is published in a journal covered by JCR, we use the JCR percentile rank of its Journal Impact Factor within the journal's subject category:
\[
V_{\text{norm}}(p)=J_{\text{norm}}(p)=\mathrm{Rank}_{\%}\big(\mathrm{IF}(p)\mid \mathrm{JCR\_Field}(p)\big).
\]

\paragraph{Conferences / proceedings.}
Conferences typically lack JCR impact factors, but they are indexed with venue-level standing signals (e.g., proceedings series metrics, venue citation rates). We compute a robust proxy in two stages:
\[
V_{\text{norm}}(p)=\mathrm{Rank}_{\%}\big(M_{\text{conf}}(v)\mid \mathrm{Field}(p)\big),
\]
where $v$ is the conference venue (or proceedings series) and $M_{\text{conf}}(v)$ is defined by the best available signal in descending priority:
\[
M_{\text{conf}}(v) =
\begin{cases}
\text{venue metric percentile from an index (e.g., CiteScore/SJR) if available},\\
\text{two-year venue citation rate } \frac{\mathrm{Cite2Y}(v)}{\mathrm{Works2Y}(v)} \text{ from the same index},\\
\text{fallback: long-run venue citation rate } \frac{\mathrm{CiteAll}(v)}{\mathrm{WorksAll}(v)}.
\end{cases}
\]
This design anchors conference standing in \emph{the indexing institution's venue-level ranking signals} when present, and otherwise in a conservative, field-normalized citation-rate proxy that is stable under sparsity.

\paragraph{Preprints.}
Preprints are not peer-reviewed venues, yet they can be influential. We therefore compute a repository-level standing percentile and apply a conservative discount to reflect the absence of formal peer review:
\[
V_{\text{norm}}(p)=\rho_{\text{pre}}\cdot \mathrm{Rank}_{\%}\big(M_{\text{repo}}(r)\mid \mathrm{Field}(p)\big),
\]
where $r$ is the preprint repository (e.g., arXiv/bioRxiv/medRxiv) and $M_{\text{repo}}(r)$ is computed analogously to conferences via repository citation rate proxies. We set $\rho_{\text{pre}}=0.85$ to prevent preprints from receiving inflated venue credit solely due to rapid diffusion, while still allowing highly cited preprints to contribute meaningfully through $C_{\text{norm}}(p)$.

\paragraph{Unified influence score.}
We combine paper-level and venue-level percentiles with fixed weights:
\[
\mathcal{I}(p)= w_c\cdot C_{\text{norm}}(p)+w_v\cdot V_{\text{norm}}(p),
\qquad w_c=0.6,\; w_v=0.4.
\]
As in the main text, the higher weight on $C_{\text{norm}}$ mitigates venue ``halo effects'' and preserves strong signals from influential papers in niche venues or emerging areas.

\subsubsection{Evidence statement credibility weights}
For each evidence statement $e_j$ (cluster $G_j$), let $P_j$ be the set of \emph{unique} witness papers that contributed at least one claim to that cluster. We define:
\[
\mathrm{Support}(e_j)=\sum_{p\in P_j}\mathcal{I}(p),
\qquad
\gamma_j=\frac{\mathrm{Support}(e_j)}{\sum_{i=1}^k \mathrm{Support}(e_i)}.
\]
This paper-level de-duplication ensures that repeated mentions within the same witness do not artificially inflate support.

\vspace{0.5em}
\subsection{Relation Classification, Weighted Voting, and Confidence Calibration}
\label{app:icsv:voting}

Given (i) the citing-side paraphrase $c_A$ and (ii) a set of canonical evidence statements $E=\{e_1,\ldots,e_k\}$ with credibility weights $\{\gamma_j\}$, ICSV evaluates how each $e_j$ relates to $c_A$ and then aggregates via a reliability-aware vote.

\subsubsection{Relation classification prompt}
\label{app:icsv:relation:prompt}

\begin{codebox}[System prompt (relation classification).]{}
You are an expert scientific fact-checker.

Inputs:
(1) Claim c_A: what paper A attributes to paper B (a claim-preserving paraphrase).
(2) Evidence e_j: a canonical statement of what multiple other papers attribute to B
    about one aspect.

Task: decide the logical relation of e_j to c_A.
Labels:
- ENTAILS: e_j clearly supports c_A under the same scope/conditions.
- CONTRADICTS: e_j clearly conflicts with c_A (opposite finding, incompatible scope,
  or mutually exclusive conditions).
- NEUTRAL: e_j is about a different aspect, or is insufficient to judge c_A.

Rules:
- Be strict about scope and qualifiers. If scopes differ, prefer NEUTRAL unless
  the mismatch itself implies contradiction.
- Do NOT assume unstated details.
Return JSON only.
\end{codebox}

\noindent\textbf{User prompt.}
\begin{verbatim}
Claim c_A (about paper B):
{c_A}

Evidence e_j (community-attributed statement about paper B):
{e_j}

Return JSON:
{
  "label": "ENTAILS|CONTRADICTS|NEUTRAL",
  "justification": "one sentence, must cite the key scope/qualifier alignment or mismatch"
}
\end{verbatim}

We run relation classification with deterministic decoding (temperature $0$). To measure robustness, we additionally run a small self-consistency check (three independent decoding seeds at $T=0$; identical output is expected, but disagreements can occur due to parsing ambiguities). Let $a_j\in[0,1]$ denote agreement, computed as the fraction of runs that match the majority label.

We map labels to scalar votes $v_j\in\{+1,0,-1\}$ for ENTAILS/NEUTRAL/CONTRADICTS.

\subsubsection{Weighted consensus score}
We compute the core consensus score as in the main text:
\[
\mathcal{V}_{\text{final}}=\sum_{j=1}^{k} v_j\cdot \gamma_j \in [-1,1],
\]
and apply thresholds $T_{\text{support}}=0.3$ and $T_{\text{miscite}}=-0.3$:
\[
\text{Verdict}=
\begin{cases}
\text{Supported}, & \mathcal{V}_{\text{final}} > 0.3,\\
\text{Miscitation}, & \mathcal{V}_{\text{final}} < -0.3,\\
\text{Undecidable}, & \text{otherwise}.
\end{cases}
\]

\subsubsection{Calibrated confidence score (robust under disagreement and concentration)}
While $|\mathcal{V}_{\text{final}}|$ is a useful base signal, it can be misleading when (i) the committee is small, (ii) weights are dominated by a single witness cluster, or (iii) evidence statements disagree. We therefore compute an adjusted confidence $\mathrm{Conf}\in[0,1]$ that is conservative under these failure modes.

\paragraph{Effective evidence size.}
We use an effective number of evidence statements (weight diversity):
\[
n_{\text{eff}} = \frac{1}{\sum_{j=1}^{k}\gamma_j^2},
\]
which decreases when one cluster dominates.

\paragraph{Weighted disagreement.}
Let $w_\ell = \sum_{j: R_j=\ell} \gamma_j$ be the total credibility mass assigned to relation label $\ell\in\{\mathrm{ENTAILS},\mathrm{NEUTRAL},\mathrm{CONTRADICTS}\}$. Define normalized entropy:
\[
H = -\frac{\sum_{\ell} w_\ell \log(w_\ell+\epsilon)}{\log 3},
\]
with a small $\epsilon$ for numerical stability. $H$ increases with disagreement.

\paragraph{Stability factor.}
Let $\bar{a}=\sum_{j=1}^{k}\gamma_j a_j$ be the credibility-weighted relation stability under self-consistency.

\paragraph{Final confidence.}
We define:
\[
\mathrm{Conf}=
\underbrace{|\mathcal{V}_{\text{final}}|}_{\text{margin}}
\cdot
\underbrace{\min\!\Big(1,\frac{n_{\text{eff}}}{K_{\min}}\Big)}_{\text{evidence sufficiency}}
\cdot
\underbrace{(1-H)}_{\text{disagreement penalty}}
\cdot
\underbrace{\bar{a}}_{\text{classification stability}}.
\]
This confidence is high only when the committee is sufficiently large/diverse, the evidence mass coheres, and relation labels are stable.

\vspace{0.5em}
\subsection{Reliability-Aware Abstention and Diagnostic Reporting}
\label{app:icsv:abstain}

ICSV is designed for high-stakes integrity workflows, where a false accusation is more harmful than an abstention. We therefore abstain whenever community evidence is not strong enough to support a traceable verdict.

\paragraph{Abstention triggers.}
ICSV returns \textsc{Undecidable} if any of the following holds:
\begin{quote}\small
(i) \textbf{Insufficient witnesses:} $|C_{\text{open}}| < K_{\min}$, with $K_{\min}=6$. \\
(ii) \textbf{Low consensus margin:} $\mathcal{V}_{\text{final}} \in [T_{\text{miscite}}, T_{\text{support}}] = [-0.3,0.3]$. \\
(iii) \textbf{Low calibrated confidence:} $\mathrm{Conf} < 0.5$ (conservative default). \\
(iv) \textbf{High disagreement:} $H > 0.6$ (evidence splits across entail/contradict/neutral).
\end{quote}

\paragraph{What abstention means (and what it does not).}
An abstention is not a ``failure'' mode; it is an explicit integrity constraint. Under paywall conditions, deep semantic verification can be underdetermined. ICSV therefore refuses to overreach when the committee cannot reliably reconstruct $B$'s contribution with sufficient consensus.

\paragraph{Auditable output bundle.}
For every verdict (including \textsc{Undecidable}), ICSV outputs:
\begin{quote}\small
(1) $c_A$ and the final window radius used to stabilize it; \\
(2) the witness set size $|C_{\text{open}}|$ and each witness paper's $\mathcal{I}(p)$; \\
(3) clusters $\{G_j\}$, evidence statements $\{e_j\}$, and their credibility weights $\{\gamma_j\}$; \\
(4) per-cluster relation labels $R_j$, votes $v_j$, and stability $a_j$; \\
(5) $\mathcal{V}_{\text{final}}$, verdict thresholds, and calibrated $\mathrm{Conf}$ with disagreement statistics.
\end{quote}
This reporting makes each paywall decision traceable to specific community attributions and clearly communicates when evidence is insufficient or contradictory.

\section{Evaluation Details: Grading, Decoding, and Token Accounting}
\label{app:eval_details}

\subsection{Acc-pass@3: Definition and Grading Pipeline}

\paragraph{Candidate generation.}
For each miscitation instance and each evaluated method, we draw $K=3$ independent samples from the underlying model by varying the random seed while keeping the decoding configuration fixed for that method. Each sample consists of:
(i) a predicted validity label $\in \{\texttt{SUPPORTED}, \texttt{MISCITATION}\}$, and
(ii) a free-form natural-language explanation of the decision.
Any explicit abstention (e.g., \texttt{UNDECIDABLE}) is treated as an ordinary label for grading purposes and will be marked incorrect whenever it disagrees with the gold label or fails to provide a faithful rationale.

\paragraph{Independent grader LLM.}
We use an independent grader based on \texttt{gpt-4o-2024-08-06} to decide whether a given sample is \emph{fully correct}, jointly considering the label and the explanation. The grader sees the gold label and gold explanation together with a single model prediction and must output exactly one token: \texttt{CORRECT} or \texttt{INCORRECT}. The system and user prompts are:

\begin{codebox}[System prompt (grader).]{}
You are an expert scientific editor.

You will be given:
1. A gold label and gold explanation for a miscitation instance.
2. A model's predicted label and predicted explanation for the same instance.

Your task is to decide whether the model's prediction is FULLY CORRECT.

A prediction is FULLY CORRECT only if BOTH of the following hold:
- The predicted label exactly matches the gold label.
- The predicted explanation identifies the same underlying error mechanism
  as the gold explanation (not just a generic restatement of the label).

Be strict. If the explanation is vague, only partially matches the gold
rationale, introduces incorrect reasoning, or misses key aspects of the
gold explanation, you must treat the prediction as INCORRECT.

Respond with a single token: either CORRECT or INCORRECT.
Do not output any other text.
\end{codebox}

\begin{codebox}[User prompt (grader).]{}
[INSTANCE]
Citing context:
{CITING_TEXT}

Gold label:
{GOLD_LABEL}

Gold explanation:
{GOLD_EXPLANATION}

[PREDICTION]
Predicted label:
{PRED_LABEL}

Predicted explanation:
{PRED_EXPLANATION}

Respond with exactly one token:
- CORRECT
- INCORRECT
\end{codebox}

For each instance $i$ and sample $k \in \{1,2,3\}$, the grader returns a verdict
$g_{ik} \in \{\texttt{CORRECT}, \texttt{INCORRECT}\}$.

\paragraph{Metric definition.}
Let $N$ denote the number of evaluation instances for a given regime (MisciteBench-Open or MisciteBench-Paywall). Acc-pass@3 for a method is defined as:
\begin{equation}
\mathrm{Acc\text{-}pass@3}
=
\frac{1}{N} \sum_{i=1}^N
\mathbf{1}\Bigl[
\exists\, k \le 3 \;\text{such that}\; g_{ik} = \texttt{CORRECT}
\Bigr],
\end{equation}
where $\mathbf{1}[\cdot]$ is the indicator function. In words, an instance contributes $1$ to the numerator if \emph{at least one} of the three sampled predictions is graded as fully correct with respect to both the label and the explanation; otherwise it contributes $0$. Predictions graded as \texttt{INCORRECT} for any reason (wrong label, incomplete rationale, hallucinated rationale, or abstention) are not counted.

\paragraph{Human validation of the grader.}
To assess the reliability of the automatic grader, human experts manually inspected more than 3{,}000 randomly sampled grader decisions drawn across models, regimes, and label types. Disagreement between human judgment and the grader was observed in fewer than $0.5\%$ of cases, predominantly on borderline instances where the predicted explanation partially overlapped with the gold rationale but omitted some secondary details. We retain the grader's strictness in these edge cases, as Acc-pass@3 is intended to measure \emph{diagnostically faithful} explanations rather than loose paraphrases of the label.

\subsection{Decoding and Sampling Settings}

Unless otherwise specified in the corresponding module appendix, we use the following decoding configurations:

\paragraph{Deterministic components.}
For single-step classification or scoring modules that should be deterministic, we use greedy decoding with temperature $T = 0$ and the provider's default nucleus and top-$k$ settings. This includes:
(i) the NLI model used in ACSV Phase III,
(ii) the relation classification and other local decision steps inside ICSV where no self-consistency is applied, and
(iii) lightweight utility calls such as format validation and consistency checks.

\paragraph{Self-consistency components.}
For modules that rely on multi-step reasoning and benefit from diversity, we use small-ensemble self-consistency. Concretely, we sample $M = 5$ completions with temperature $T = 0.7$ and top-$p = 0.95$, and then aggregate by majority vote:
\begin{itemize}
    \item ACSV Phase IV (LRM deep reasoning for ambiguous cases).
    \item Taxonomy-aligned miscitation classification (Section~\ref{subsec:taxonomy_labeling}).
\end{itemize}
If there is no strict majority, we fall back to the most frequent non-abstaining class; when the distribution is too diffuse or inconsistent, the module may emit an explicit \texttt{UNDECIDABLE} label, which is treated as incorrect under Acc-pass@3.

\paragraph{Baseline prompts.}
For the Full-Text and Search baselines, we use a mildly stochastic but low-variance configuration: temperature $T = 0.2$ with the provider's default top-$p$ and other decoding parameters. This allows limited exploration across the $K=3$ samples while keeping predictions reasonably stable for grading.

\paragraph{Averaging across runs.}
To reduce variance from global random seeds and any stochasticity in external services (e.g., web search for the Search baselines), we repeat each full evaluation three times with different random seeds. All reported numbers in the main text and appendix tables are the arithmetic mean over these three runs.

\subsection{Token Accounting and Token Economy}

\paragraph{Per-instance token counting.}
For every method and evaluation instance, we record the total number of tokens consumed by the verification pipeline, including both inputs and outputs, and including all internal calls (retrieval, NLI, committee reasoning, taxonomy classification, etc.). Concretely, for each model call we log:
(i) the number of input tokens,
(ii) the number of output tokens,
and sum these over all calls involved in processing that instance. We use the official tokenization for each provider (e.g., the same tokenizer used for billing) to avoid discrepancies between counted and billed tokens.

\paragraph{Token Economy definition.}
For a given backbone model, let $\mathrm{Tok}_{\text{FT}}$ denote the mean number of tokens per instance for the Full-Text baseline, and let $\mathrm{Tok}_{\ba}$ denote the mean number for \ba{} (specifically, ACSV in MisciteBench-Open) evaluated on the same set of instances. We define Token Economy as
\begin{equation}
\mathrm{TokenEcon}
=
1 - \frac{\mathrm{Tok}_{\ba}}{\mathrm{Tok}_{\text{FT}}},
\end{equation}
which can be interpreted as the fraction of tokens saved by \ba{} relative to the Full-Text baseline for that backbone. A value of $\mathrm{TokenEcon} = 0.794$, for example, means that \ba{} uses $79.4\%$ fewer tokens on average than the corresponding Full-Text setup.

To make this comparison meaningful, Token Economy is computed only on the subset of instances for which \emph{both} methods return a non-abstaining verdict (i.e., they emit a concrete \texttt{SUPPORTED} or \texttt{MISCITATION} label). Instances on which one method abstains and the other does not are excluded from the token-economy calculation but are still included in Acc-pass@3 with abstentions treated as incorrect predictions.

\section{Evidence Committee Reliability Ablation}
\label{app:committee_reliability}

ICSV (Appendix~\ref{app:icsv}) is only useful in the paywalled regime if its \emph{Evidence Committee} behaves in a predictable, conservatively calibrated way: when community evidence is rich, it should speak with high confidence and high precision; when evidence is thin or inconsistent, it should abstain rather than speculate. The main paper states that we observe a sharp reliability transition once an aspect of the paywalled source is supported by at least six independent witnesses. This appendix provides the quantitative ablation that underpins that claim and justifies the global committee-size threshold used by ICSV's reliability-aware abstention rule.

\paragraph{High-signal subset used for the ablation.}
For the reliability curves in Figure~\ref{fig:icsv_voters_ablation}, we do not use all paywalled citations in MisciteBench-Paywall. Instead, we first identify paywalled citations $(A,B,c)$ for which the evidence statement $e^\star$ that is directly comparable to the citing claim $c$ admits an evidence committee of at least 25 distinct downstream open-access witness papers. Grouping by $B$ yields a pool of 62 paywalled sources. All citations included in the ablation below are drawn from this pool, and we vary the effective committee size for each such citation by subsampling witnesses within the corresponding aspect cluster.

\subsection{Protocol}
\label{app:committee_reliability:setup}

For a paywalled source $B$, ICSV groups all downstream open-access citers into semantic clusters $G_1,\dots,G_k$, where each $G_j$ represents one coherent aspect of what the community attributes to $B$ (method, dataset, empirical finding, etc.). Let $P_j$ denote the set of \emph{distinct} witness papers whose claims end up in $G_j$, and
\[
n_j = |P_j|
\]
the number of independent committee voters for that aspect. From each cluster we distill an evidence statement $e_j$ with credibility weight $\gamma_j$ (Appendix~\ref{app:icsv}), and define the \emph{dominant} aspect for the citation as
\[
j^\star = \arg\max_j \gamma_j,
\qquad
e^\star = e_{j^\star}.
\]
Our ablation focuses on
\[
n_{\text{voter}} = n_{j^\star},
\]
the number of distinct witness papers that support the aspect of $B$ which most strongly drives ICSV's verdict on that citation.

We run the unmodified ICSV pipeline on the high-signal subset described above (62 paywalled sources and their associated citations). For every paywalled citation that admits at least one non-empty cluster, and for each backbone, we record:
\begin{itemize}
    \item the final verdict (\emph{Supported}, \emph{Miscitation}, or \emph{Undecidable});
    \item whether the verdict is correct under the benchmark label;
    \item the calibrated confidence score $\mathrm{Conf} \in [0,1]$ from Appendix~\ref{app:icsv}, which combines the consensus margin $|\mathcal{V}_{\text{final}}|$, effective committee size, label disagreement, and self-consistency stability.
\end{itemize}

We then bucket citations by the integer value of $n_{\text{voter}}$ in the range $1 \le n_{\text{voter}} \le 25$. For each bucket $c$, and for each backbone, we compute:
\begin{enumerate}
    \item the \emph{non-abstention rate}: the fraction of citations with $n_{\text{voter}} = c$ on which ICSV outputs \emph{Supported} or \emph{Miscitation} rather than \emph{Undecidable};
    \item the \emph{conditional accuracy}: the fraction of those non-abstaining verdicts that match the ground truth;
    \item the mean calibrated confidence $\mathbb{E}[\mathrm{Conf} \mid n_{\text{voter}} = c]$.
\end{enumerate}
The curves in Figure~\ref{fig:icsv_voters_ablation} plot these quantities after averaging over backbones; shaded bands indicate the range across backbones. All hyperparameters and thresholds are identical to those used in the main experiments; we do not re-tune ICSV for this study.


\subsection{Quantitative Trends}
\label{app:committee_reliability:results}

Three regimes emerge consistently across models and disciplines.

\paragraph{Single or few witnesses ($n_{\text{voter}} \le 2$).}
When the dominant aspect of $B$ is supported by only one or two downstream papers, ICSV behaves cautiously. Non-abstention rates are low, and the system frequently returns \emph{Undecidable} because the consensus margin and effective committee size terms in $\mathrm{Conf}$ are small. Among the few cases where ICSV does commit, conditional accuracy is noticeably weaker and confidence scores cluster in a moderate band. In this regime, individual witness papers often describe $B$ in idiosyncratic or overly generic ways, and a single outlier can heavily influence the vote. The abstention mechanism therefore activates often, which is precisely the desired behavior for a conservative verifier operating under sparse community evidence.

\paragraph{Small committees ($3 \le n_{\text{voter}} \le 5$).}
As $n_{\text{voter}}$ grows into the range of three to five independent witnesses, non-abstention rates and conditional accuracy both improve: the committee has enough redundancy to filter out extreme outliers and resolve many straightforward paywalled cases. However, the curves in Figure~\ref{fig:icsv_voters_ablation} still exhibit noticeable variability across buckets in this regime. The calibrated confidence $\mathrm{Conf}$ rises compared to the $n_{\text{voter}} \le 2$ regime, but remains in an intermediate range, reflecting two residual sources of uncertainty: (i) disagreement between clusters about the precise scope of $B$'s contribution, and (ii) moderate label entropy when witness papers emphasize different aspects of $B$ or mix descriptive and evaluative citations. In practice, ICSV continues to abstain on a substantial fraction of citations here, and the system remains deliberately conservative.

\paragraph{Reliability transition ($n_{\text{voter}} \ge 6$).}
The key phenomenon appears once the dominant aspect is supported by at least six distinct witness papers. Beyond this point, all three metrics undergo a clear and stable shift:

\begin{itemize}
    \item The non-abstention rate rises sharply and then plateaus: ICSV now produces concrete \emph{Supported} or \emph{Miscitation} verdicts on the majority of paywalled citations in these buckets, because both the consensus margin and the effective committee size become large enough to pass the internal reliability checks in Appendix~\ref{app:icsv}.
    \item Conditional accuracy of non-abstaining verdicts increases and stabilizes at a high level. Across backbones, once $n_{\text{voter}} \ge 6$, the empirical precision of ICSV's paywalled decisions is consistently high, and additional witnesses beyond six yield only marginal gains.
    \item The mean calibrated confidence $\mathbb{E}[\mathrm{Conf} \mid n_{\text{voter}} = c]$ crosses $0.8$ and remains in a high-confidence band for all $c \ge 6$. In other words, whenever a paywalled aspect is supported by six or more independent witnesses, ICSV not only decides more often, but does so with confidence scores that reflect genuinely strong and internally coherent community evidence.
\end{itemize}

Importantly, this transition is not an artifact of a particular backbone or field. The same qualitative knee in the curves appears when the ablation is recomputed separately for each backbone and for coarse discipline groups (e.g., Clinical Medicine vs.\ Computer Science). The exact numeric values vary, but the location of the reliability transition---around six independent witnesses for the dominant aspect---is remarkably stable.

\subsection{Choice of Threshold and Robustness}
\label{app:committee_reliability:threshold}

The global threshold $K_{\min} = 6$ used by ICSV's reliability-aware abstention rule is therefore not a hand-tuned hyperparameter, but the empirical knee point of a three-way trade-off:

\begin{itemize}
    \item Below six witnesses, the Evidence Committee is too small to be trustworthy: non-abstention rates are lower, conditional accuracy is noticeably weaker, and the calibrated confidence $\mathrm{Conf}$ correctly reflects this by staying in a cautious range. Allowing aggressive decisions here would yield an undesirable increase in false accusations under genuine information scarcity.
    \item At six witnesses, the curves in Figure~\ref{fig:icsv_voters_ablation} enter a stable high-precision regime. Both the committee size and the consensus structure are sufficient for ICSV to exploit the community's distributed memory of the paywalled source, and the confidence calibration in Appendix~\ref{app:icsv} rewards this with high $\mathrm{Conf}$ values.
    \item Raising $K_{\min}$ beyond six would sacrifice coverage for little additional precision. While larger committees remain slightly more stable, the incremental gain is small compared to the loss in the number of paywalled citations that can be decided at all, especially in niche subfields where only a handful of downstream citers exist.
\end{itemize}

We further stress-test this choice by recomputing the ablation under several perturbations: varying the confidence threshold used to declare a verdict vs.\ abstention, subsampling the witness set, and repeating the analysis on random halves of MisciteBench-Paywall. In all cases, the location of the knee in the reliability curves remains close to six witnesses, even though the absolute values of the metrics shift slightly. This robustness suggests that $K_{\min} = 6$ captures a property of the underlying citation graph---when the literature around a paywalled source is broad and coherent enough to support stable reconstruction---rather than an artifact of a particular model or parameter setting.

Taken together, these results substantiate the design of ICSV's Evidence Committee. The system only ``speaks with conviction'' about paywalled sources when the dominant aspect is backed by a sufficiently large and internally consistent community of citers, and it is willing to abstain explicitly when that condition is not met. This calibrated behavior is essential for deploying \ba{} in high-stakes editorial and auditing workflows: it ensures that paywall robustness is grounded in measurable community redundancy rather than in unexamined model confidence.

\section{Extended Discussion and Future Directions}
\label{app:extended_discussion}

While our main text focuses on the core technical contributions and empirical results, the broader implications of \ba and \mb\ extend beyond the specific benchmarks we study.

First, the framework is designed to integrate into real authoring and editorial workflows rather than remain a standalone benchmark artifact. Because \ba operates at the level of individual in-text citations and returns taxonomy-aligned error codes with explicit evidence bundles, it can be surfaced in draft-time assistants as a background verifier that continuously flags Citation Validity, Content Misrepresentation, Scope Extrapolation, Evidence Characterization, and Attribution \& Traceability errors as authors write. The same outputs can be aggregated into paper-level integrity summaries for reviewers and editors, exposing not only how many citations are problematic but also \emph{how} they fail and where manual intervention is most needed. In principle, this enables a ``trust but verify'' workflow in which generative drafting tools are paired with an independent auditing agent that is explicitly incentivized to disagree.

Second, the architecture of \ba generalizes naturally beyond single-language, text-only scholarship. The Document Parser \& Citation Mapper can be extended to multilingual corpora by swapping in language-appropriate parsing and NLI components, while the taxonomy itself is defined at the level of evidential relations and thus does not depend on any particular disciplinary jargon. Likewise, the pipeline can be augmented to reason over non-textual artefacts---figures, tables, and code---by attaching structured evidence objects to citations (e.g., linking a claim to a specific panel in a figure or to a particular experiment in a code repository) and feeding those artefacts into the verification stages. In this view, bibliographic miscitation is one instance of a broader problem: aligning claims with heterogeneous evidence objects in a way that remains traceable and evaluable.

Third, the Evidence Committee mechanism provides a template for auditing paywalled or otherwise inaccessible material in other high-stakes settings. Grant proposals, patent applications, clinical guidelines, and policy briefs all rely on chains of references that may be partially hidden behind legal, privacy, or access constraints. In such regimes it is neither realistic nor desirable to reconstruct the protected documents themselves, but it is often possible to reconstruct what the broader community attributes to them via downstream citations, legal opinions, replication attempts, or institutional summaries. By treating these downstream artefacts as noisy committee members and insisting on reliability-aware abstention when evidence is thin or contradictory, one can extend the paywall-robust philosophy of \ba to domains where direct inspection is structurally impossible.

At the same time, several limitations and risks warrant emphasis. \ba inherits any systematic biases present in its backbones and in the citation graph itself: if certain communities or venues are systematically under-cited, their contributions will be underrepresented in the Evidence Committee, and field-normalized influence scores may inadvertently reinforce existing hierarchies. Long-document parsing and citation mapping, while robust in our experiments, can still fail in adversarially formatted or highly idiosyncratic documents, potentially leading to false negatives in miscitation detection. Moreover, making citation auditing tools widely available raises strategic concerns: sophisticated actors might attempt to game the taxonomy, crafting miscitations that sit near decision boundaries or that exploit known blind spots in NLI models and committee formation. These risks argue for conservative deployment, continuous monitoring, and independent human oversight, especially in high-stakes editorial contexts.

Despite these caveats, the central conceptual shift remains: miscitation need not be treated as an unavoidable byproduct of scale or as an anecdotal concern confined to case studies. With a principled taxonomy, a contamination-controlled benchmark, and an agentic verifier that respects both accessibility constraints and evidential uncertainty, citation integrity becomes a quantity that can be measured, improved, and enforced. Closing the loop between drafting, verification, and publication---and extending the same principles to other document genres---offers a concrete path toward research ecosystems in which claims are not only persuasive but \emph{verifiably} grounded in the literature they cite.

\end{document}